\documentclass[lettersize,journal]{IEEEtran}
\usepackage{amsmath,amssymb,amsfonts}
\usepackage{algorithm}
\usepackage{algorithmicx}
\usepackage{array}
\usepackage[caption=false,font=normalsize,labelfont=rm,textfont=rm]{subfig}
\usepackage{textcomp}
\usepackage{stfloats}
\usepackage{url}
\usepackage{verbatim}
\usepackage{graphicx}
\usepackage{epstopdf}
\usepackage{cite}
\usepackage{algpseudocode}
\usepackage{subfloat} % subfigure
\usepackage{caption}
\usepackage{hyperref}
\usepackage{float}
\usepackage{subfig}
\usepackage{bm}
\usepackage{gensymb}
\usepackage{multirow}
\usepackage{array}
\begin{document}

\title{Hybrid Near/Far-Field Frequency-Dependent Beamforming via Joint Phase-Time Arrays
}

\author{Yeyue Cai, Meixia Tao, Jianhua Mo, Shu Sun
        % <-this % stops a space ,~\IEEEmembership{Staff,~IEEE,}
\thanks{The authors are with the Department of Electronic Engineering and the Cooperative Medianet Innovation Center (CMIC), Shanghai Jiao Tong University, China (email:caiyeyue@sjtu.edu.cn, mxtao@sjtu.edu.cn, mojianhua01@gmail.com, shusun@sjtu.edu.cn).\par
Part of this work was accepted by IEEE Wireless Communications and Networking Conference (WCNC), Milan, Italy, March 2025 \cite{wcnc}.
}% <-this % stops a space
}

\markboth{ }%
{Shell \MakeLowercase{\textit{et al.}}: A Sample Article Using IEEEtran.cls for IEEE Journals}

\maketitle

\begin{abstract}
Joint phase-time arrays (JPTA) emerge as a cost-effective and energy-efficient architecture for frequency-dependent beamforming in wideband communications by utilizing both true-time delay units and phase shifters. This paper exploits the potential of JPTA to simultaneously serve multiple users in both near- and far-field regions with a single radio frequency chain. The goal is to jointly optimize JPTA-based beamforming and subband allocation to maximize overall system performance. To this end, we formulate a system utility maximization problem, including sum-rate maximization and proportional fairness as special cases. We develop a 3-step alternating optimization (AO) algorithm and an efficient deep learning (DL) method for this problem. The DL approach includes a 2-layer convolutional neural network, a 3-layer graph attention network (GAT), and a normalization module for resource and beamforming optimization. The GAT efficiently captures the interactions between resource allocation and analog beamformers. Simulation results confirm that JPTA outperforms conventional phased arrays (PA) in enhancing user rate and strikes a good balance between PA and fully-digital approach in energy efficiency. Employing a logarithmic utility function for user rates ensures greater fairness than maximizing sum-rates. Furthermore, the DL network achieves comparable performance to the AO approach, while having orders of magnitude lower computational complexity.
\end{abstract}


\section{Introduction}
To meet the stringent demands for high data rates, ultra-low latency, and hyper-reliability, the sixth-generation (6G) communication system is anticipated to shift to higher frequency bands, including millimeter-wave and terahertz (THz) \cite{wang2023road}. However, these higher frequency bands experience severe propagation attenuation due to increased carrier frequencies \cite{rappaport2019wireless}. Extremely large-scale multiple-input multiple-output (XL-MIMO) techniques have been proposed as a promising solution to mitigate signal attenuation in high-frequency bands by focusing radio waves towards desired areas through beamforming \cite{busari2017millimeter,bjornson2019massive}. While the integration of XL-MIMO and high-frequency technologies is seen as crucial for 6G, it also presents new challenges. 

One primary challenge lies at the efficient XL-MIMO beamforming architecture, specially over wideband communication channels. In traditional fully-digital (FD) beamforming architecture, each antenna element is required to connect with a dedicated radio frequency (RF) chain. This becomes highly impractical in XL-MIMO systems due to the prohibitive hardware costs and power consumption. Hybrid beamforming (HBF) architectures by using analog devices, such as phase shifters (PSs) and switches, are cost-effective alternatives \cite{molisch2017hybrid}. However, PS-based HBF can only produce frequency-flat spatial responses and fails to align with the frequency-dependent variations in array response of the wideband communication channel, thus causing severe spatial-wideband effect \cite{el2014spatially}. Recently, joint phase-time arrays (JPTA), integrating true-time delays (TTDs) to provide programmable phase adjustments across frequencies, has been introduced as an innovative solution to these challenges \cite{rotman2016true}. By dynamically adjusting phase on each subband, JPTA significantly enhances the flexibility and directionality of beams, thereby mitigating the spatial-wideband effect and promoting more efficient spectrum utilization. Nevertheless, the full potential of JPTA architecture remains largely unexplored.

Another major challenge faced by XL-MIMO over high-frequency bands is the significant extension of the Rayleigh distance, the boundary between near-field (NF) and far-field (FF) regions. In the traditional FF region, the wireless channel is modeled under the planar wave assumption and, therefore, FF beamforming mainly targets specific angles. By contrast, the wavefront in the NF propagation is spherical and thus NF beamforming should consider both angle and distance \cite{liu2023near1,liu2023near}. The significant extension of Rayleigh distance requires precise beam focusing at specific locations where traditional FF beamforming is no longer suitable.  \cite{zhang20236g}. 

The aim of this work is to explore the potential of JPTA architecture to generate frequency-dependent beamformers for hybird near-far field communications.

\subsection{Related Works}
\subsubsection{TTD-Based HBF}
Recent years have witnessed significant advancements in TTD-based beamforming technologies, for both FF \cite{dai2022delay,ratnam2022joint,yildiz20243d,alammouri2022extending,jain2023mmflexible,zhao2024fast,10179244,gao2023integrated} and NF \cite{10271123,cui2022near,10541333,guo2024wideband,10458958,ting2024adaptive,zhang2023deep} communications. From a system perspective, the utilization of TTDs is broadly categorized into three areas: spreading different frequency bands across various user directions simultaneously with a single RF chain \cite{ratnam2022joint,yildiz20243d,alammouri2022extending,jain2023mmflexible,zhao2024fast}, reducing beam training costs with controllable rainbow beams \cite{10179244,cui2022near,gao2023integrated,10271123}, and mitigating the spatial-wideband effect with unidirectional beams \cite{10541333,guo2024wideband,10458958,ting2024adaptive}. More specifically, in the context of generating multiple frequency-dependent beams for multiple users, studies \cite{ratnam2022joint,yildiz20243d,alammouri2022extending,jain2023mmflexible,zhao2024fast} have demonstrated the effectiveness of a TTD-based architecture with a single RF chain, called JPTA. This architecture has been shown to improve array gain \cite{ratnam2022joint,yildiz20243d}, boost uplink throughput while reducing latency \cite{alammouri2022extending}, achieve higher spectrum usage \cite{jain2023mmflexible}, and improve spectral efficiency (SE) \cite{zhao2024fast}. For rainbow beam generation, TTD-based beamforming has proven effective in achieving higher data rates with reduced overhead in beam alignment and enhancing the precision of integrated sensing and communication systems by significantly reducing the angle and distance sensing errors in both NF and FF scenarios \cite{10179244,cui2022near,gao2023integrated,10271123}. In tackling the spatial-wideband effect, integrating TTD components into beamforming structures has been shown to not only improve array gain but also enhance achievable rates with both uniform linear array (ULA) \cite{10541333} and uniform circular array \cite{guo2024wideband}. Furthermore, various TTD configurations have been explored to assess their impact across different NF communication scenarios in \cite{10458958} and \cite{ting2024adaptive}, demonstrating their adaptability and effectiveness in handling spatial-wideband effect.

\subsubsection{Learning-Based Beamforming Design}
Recently, deep learning (DL) has emerged as a promising approach to reduce computational complexity while maintaining or surpassing the performance of traditional algorithms. Various studies have explored DL applications in optimizing HBF problems. In \cite{lin2019beamforming}, a neural network is proposed to solve a sum-rate maximization problem. The work \cite{liu2022deep} treats antenna selection as a classification task and jointly optimizes the antenna selection and HBF design through a two-layer convolutional neural network (CNN). The capabilities of TTD-based HBF architectures in combating spatial-wideband effects in THz NF communication scenarios using DL methods are explored in \cite{ting2024adaptive} and \cite{zhang2023deep}. To be specific, the work \cite{ting2024adaptive} combines a U-net and a transformer network to optimize hybrid beamforming with adaptively connected TTDs for a wideband multi-user scenario. The work \cite{zhang2023deep} decomposes the time delays and phase shifts design problem into two subproblems and proposes a fully-connected (FC) neural network and a low-complexity geometry-assisted method to configure the beamforming design. Moving beyond traditional DL architectures, graph neural networks (GNNs) have been adopted to leverage the graph-structured topologies of devices and beamforming structures, enhancing the generalization and scalability of DL methods. The work in \cite{shen2022graph} demonstrates the potential of GNNs to efficiently scale to larger networks without requiring pretraining when addressing sum-rate maximization in device-to-device networks. Furthermore, graph attention networks (GATs) have been employed for beamforming optimization in \cite{li2024gnn}. This approach leverages attention-enabled aggregation and a residual-assisted combination strategy to more effectively capture user associations, yielding superior results compared to conventional GNN methods.

\subsection{Contributions}
As mentioned above, several different use cases of TTDs have been investigated in wideband systems. However, the potential of JPTA to serve multiple users simultaneously in NF or hybrid field through frequency-dependent beams remains unexplored. Furthermore, existing approaches to JPTA-based beamforming problems typically rely on iterative or heuristic algorithms. The use of DL, which offers a low-complexity and feasible real-time application solution, has not yet been explored in the context of frequency-dependent beamforming design with JPTA.
% which may not be suitable for real-time applications due to the high computational complexity \cite{yildiz20243d, zhao2024fast}

To fill the above gaps, this paper explores a wideband system where a base station (BS) equipped with a JPTA and a single RF chain serves multiple NF and FF users. The integration of NF introduces significant challenges for resource allocation and beamforming strategies due to the complex dependencies of channel characteristics on both angle and distance, which causes the beam focusing function. This complexity indicates that users in the same direction may experience different channel characteristics, significantly complicating beamforming challenges beyond those in traditional FF scenarios. Additionally, variability in user distances within the hybrid fields results in non-uniform channel gains across different links, necessitating a meticulously designed resource allocation strategy to ensure equitable service quality across diverse user locations. To tackle these challenges, we first propose a 3-step alternating optimization (AO) algorithm. Furthermore, to offer a real-time and low-complexity alternative, we introduce an innovative DL approach based on the GAT \cite{veličković2018graph}, which directly learns the resource allocation and beamforming vectors from statistical channel information. The key contributions of this work are as follows:

\begin{itemize}
    \item Leveraging JPTA's beam-splitting effects, we introduce a method to simultaneously serve multiple NF and FF users using just a single RF chain over a wideband channel, ensuring that each user receives high beamforming gain within their allocated frequency bands.
    \item We address the complex frequency subband allocation and JPTA-based beamforming by formulating a network utility maximization problem. The utility of each user is a concave, increasing, and continuously differentiable function of the transmission rate, and it is used to strike a balance between the overall system throughput and max-min fairness among users.
    \item We propose a 3-step AO algorithm to solve the formulated problem, which is a mixed integer non-linear programming (MINLP) problem. We also provide a detailed analysis of its computational complexity. 
    \item Additionally, we propose a novel, efficient algorithm based on a DL network. This network integrates a 2-layer CNN with a 3-layer node-wise GAT and a normalization module. The proposed node-wise GAT can efficiently capture the interactions between resource allocation and analog beamformers and assign dynamic weights for different graph nodes. 
    % Numerical results demonstrate that the unsupervised learning approach achieves performance comparable to the AO algorithm, but with lower computational complexity, which is more efficient for real-time use.
    \item Our numerical results validate the effectiveness of the JPTA architecture, showing that it can provide higher antenna gains at specified frequency bands. Compared to conventional phased arrays (PA), the JPTA achieves user rates that are 8.21\% and 8.07\% higher in the 2-user and 5-user scenarios, respectively. Furthermore, results indicate that the logarithmic utility function outperforms the linear utility function in balancing user fairness, offering a more equitable distribution of resources across users with varying locations. Besides, JPTA can strike a good balance between FD and PA in the term of energy efficiency (EE).
\end{itemize}

\subsection{Paper Organization and Notations}
The remainder of this paper is organized as follows. In Section \ref{sec: system model}, the signal model with the considered JPTA architecture and signal model are described. Section \ref{sec: AO} investigates the resource allocation and beamforming design through the 3-step AO method. Section \ref{sec: DL} proposes a node-wise GAT-based learning approach to solve the optimization problem in an unsupervised way. Section \ref{sec: results} outlines the numerical results of different array structures and optimization methods. Finally, conclusions are drawn in Section \ref{sec: conclusion}. 

\textit{Notations:} The transpose and conjugate transpose of a matrix are denoted by \([\cdot]^T\) and \([\cdot]^H\), respectively. The Euclidean norm of a vector is represented by \(||\cdot||\). We define \(\left[ 
\mathbf{A} \right]_{i,j}\) and \(\left[ 
\mathbf{A} \right]_{i,:}\) as the \((i,j)\)-th element and \(i\)-th row of matrix \(\mathbf{A}\). A block diagonal matrix with diagonal blocks \(\mathbf{a}_1,\ldots, \mathbf{a}_N\) is denoted as \(\operatorname{blkdiag}\left\{{\mathbf{a}}_{1}, \ldots, {\mathbf{a}}_{N}\right\} \). The operation \(\mathbf{A} \oplus_i \mathbf{B} \) denotes the concatenation of matrices \(\mathbf{A}\) and  \(\mathbf{B}\) along the \(i\)-th dimension. \(\left[N\right]\) denotes the set of integers \(\{1,2,\ldots, N\}\).

\begin{figure*}[t]
    \centering    
    \includegraphics[width=0.85\linewidth]{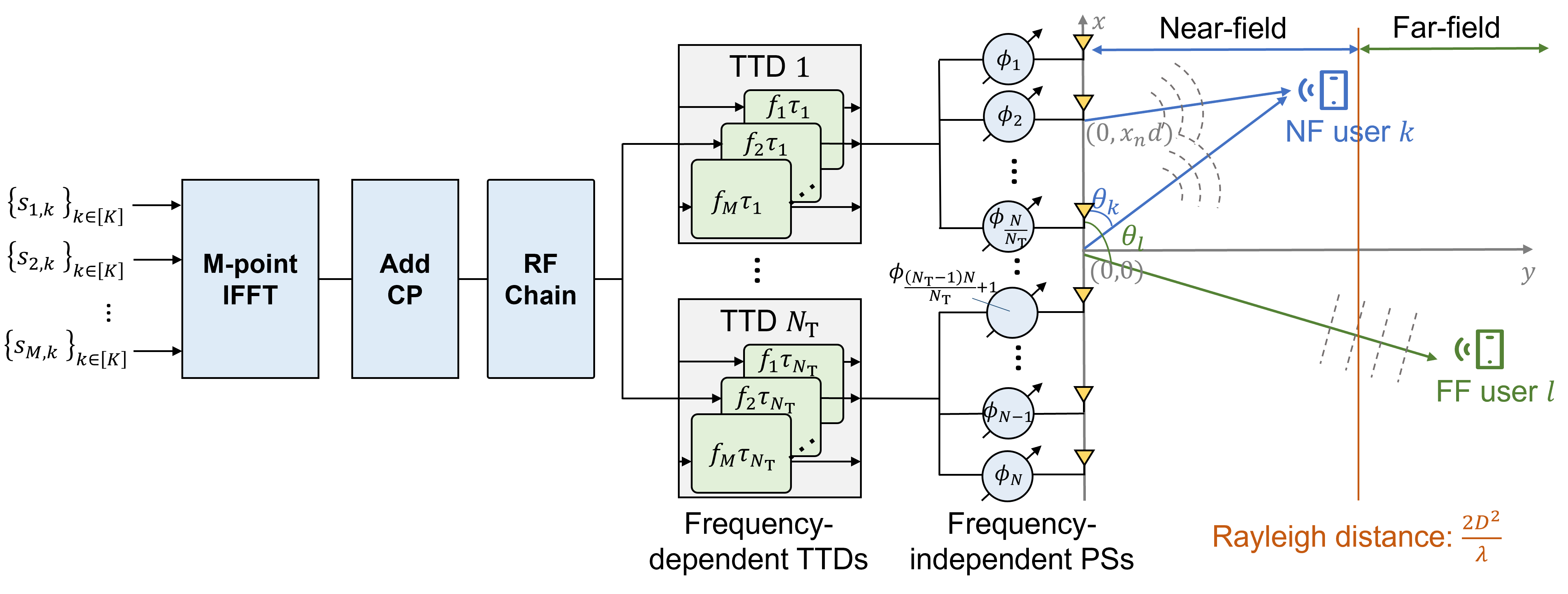}
    \caption{The JPTA with a single RF chain, $N_{\rm{T}}$ TTDs and $N$ PSs in a hybrid near-far field OFDM communication system.}
    \label{fig: system model}
\end{figure*}

\section{System Model and Problem Formulation}
\label{sec: system model}
\subsection{System Model}
We consider a hybrid near-far field orthogonal frequency division multiplexing (OFDM) wideband multi-user communication system, where the BS is equipped with a single RF chain and a large-scale ULA of $N$ elements. As shown in Fig. \ref{fig: system model}, we adopt the JPTA architecture to facilitate frequency-dependent beamforming, where the single RF chain is connected to \(N_{\rm{T}}\) TTDs and each TTD is further connected to a dedicated sub-antenna array through PSs. This system concurrently serves $K$ single-antenna users, among which the first $K_{\mathrm{NF}}$ users are located in the NF region and the remaining $K_{\mathrm{FF}}\!=\!K\!-\!K_{\mathrm{NF}}$ users in the FF region. We adopt the Rayleigh distance, defined as $r_{\rm{Rayleigh}}=\frac{2D^2}{\lambda}$, where $D$ and $\lambda$ denote the antenna array aperture and carrier wavelength respectively, to distinguish between the NF and FF regions. There are $M$ orthogonal subbands to be allocated, with each subband consisting of multiple adjacent subcarriers. It is assumed that all channels within each subband is frequency-flat. Let $B$ denote the system bandwidth and $f_c$ the central carrier frequency. Accordingly, subband \(m\) has a center frequency of $f_m\!=\!f_c+\frac{B(2m-1-M)}{2M}, \forall m \in\left[M\right] $.

\subsubsection{Channel Models for Near and Far Fields}
We primarily focus on the dominant line-of-sight component for both NF and FF users. The position of user \(k\) relative to the center of the ULA at the BS is characterized by the angle $\theta_{k}$ and distance $r_{k}$. The coordinates of user \(k\) and the $n$-th element of the ULA are denoted by $\mathbf{u}_{k}=\left[r_{k} \cos \theta_{k}, r_{k} \sin \theta_{k}\right]^{T}$ and $\mathbf{c}_{n}=\left[x_{n} d, 0\right]^{\mathrm{T}}, \forall n \in\left[N\right]$, where $x_{n}=n-\frac{N+1}{2}$ and $d=\frac{c}{2f_c}$. 
%  non-uniform spherical wavefront propagation model

In the NF region, the propagation distance from the $n$-th antenna element to user $k$ ($\forall k \in\{1,\ldots,K_{\rm{NF}}\}$) is approximated following the spherical wave model as \cite{liu2024near} 
\begin{equation}
r_{k, n}=\left\|\boldsymbol{u}_{k}-\boldsymbol{c}_{n}\right\|{\approx} r_{k}-x_{n} d \cos \theta_k +\frac{x_{n}^2 d^2 \sin^2{\theta_k}}{2r_k}.
\end{equation}

For far-field conditions (\(r_k > r_{\rm{Rayleigh}}\)), where the planar wavefront is assumed, the propagation distance simplifies to
\begin{equation}
r_{k, n} {\approx} r_{k}-x_{n} d \cos \theta_k,
\forall k \in\{K_{\rm{NF}}+1,\ldots,K_{\rm{NF}}+K_{\rm{FF}}\}.
\end{equation}

The channel for subband \(m\) of user $k$ can be modeled as 
\begin{equation}
\mathbf{h}_{m, k}=\tilde{\beta}_{m, k} \mathbf{a}_{m,k} ,\forall m\in\left[M\right],k\in\left[K\right],
\end{equation}
where $\tilde{\beta}_{m, k}=\beta_{k} e^{-j \frac{2 \pi f_{m}}{c} r_{k}}$, and $\beta_{k}=\frac{c}{4\pi f_mr_k}$ denote the complex channel gain. Vector $\mathbf{a}_{m,k}$ denotes the array response vector, which is given by
\begin{equation}
\mathbf{a}_{m,k} \!=\! \left[e^{-j \frac{2 \pi f_m}{c}\left(r_{k, 1}-r_{k}\right)}, \ldots, e^{-j \frac{2 \pi f_m}{c}\left(r_{k, N}-r_{k}\right)}\right]^{T}.
\end{equation}

\subsubsection{Signal Model with JPTA beamforming}
Define the binary subband allocation variable $b_{m,k}\in \left\{0,1\right\}$ to indicate whether subband \(m\) is assigned to user \(k\), with $b_{m,k}=1$ indicating that subband \(m\) is allocated to user \(k\), and $b_{m,k}=0$ otherwise. The received signal of user $k$ can be modeled as:
\begin{equation}
\label{eq:y}
{y}_{m,k}
=b_{m,k}\sqrt{p_m}\mathbf{h}_{m,k}^H \mathbf{\Phi} \mathbf{T}_{m}  s_{m,k} + n_{m, k},
\end{equation}
where $s_{m,k}$ is the information data for user $k$ on subband $m$ and $n_{m,k}$ represents the zero-mean additive complex Gaussian white noise with variance of $\sigma^2$. The term $p_{m}$ denotes the power allocated to subband $m$. The matrices $\mathbf{\Phi} \in \mathbb{C}^{N \times N_{\rm{T}}}$ and $\mathbf{T}_{m} \in \mathbb{C}^{ N_{\rm{T}} \times1}$ correspond to the frequency-independent and frequency-dependent analog beamformers, implemented by PSs and TTDs, respectively. 

In the sub-connected array configuration, the PS-based analog beamformer $\mathbf{\Phi}$ is given as 
\begin{equation}
\mathbf{\Phi}=\frac{1}{\sqrt{N}}\operatorname{blkdiag}\left\{{\bm{\phi}}_{1}, \ldots, {\bm{\phi}}_{N_{\rm{T}}}\right\} ,
\end{equation}
where ${\bm{\phi}}_{i} \in \mathbb{C}^{\frac{N}{N_{\rm{T}}} \times 1}$ denotes the PS-based beamformer for the subarray connected to the $i$-th TTD, with the unimodular constraint $\left| \left[ {\bm{\phi}}_{i} \right] _j \right| =1 $ for each entry $j \in \left[\frac{N}{N_{\rm{T}}}\right]$. 

The TTD-based analog beamformer is given by
\begin{equation}
\label{eq:TTD1}
\mathbf{T}_{m}=\left[e^{-j 2 \pi f_{m} \tau_{1}}, \ldots, e^{-j 2 \pi f_{m} \tau_{N_{\rm{T}}}}\right]^{T},
\end{equation}
where $\tau_{i}$, for $ i \in \left[N_{\rm{T}}\right]$, denotes the time delay imparted by the $i$-th TTD within the range $\left[0, \tau_{\max }\right]$.

Then, the array gain realized by PSs and TTDs on an arbitrary physical location \((\theta,r)\) at frequency \(f_m\) can be written as
\begin{equation}
    G(f_m,\theta,r) = \left|  \mathbf{a}_{m,k}^H\mathbf{\Phi} \mathbf{T}_{m} \right|^2.
\end{equation}

\subsection{Problem Formulation}
In this study, we aim to maximize the sum of utility functions of downlink transmission rates of all users by optimizing the subband allocation variable \(\{b_{m,k}\}_{m\in \left[M\right],k\in \left[K\right]}\), power allocation variable \(\{p_m\}_{m\in \left[M\right]}\), PS-based analog beamformers \(\{{\bm{\phi}}_{i}\}_{i\in \left[N_{\rm{T}}\right]}\), and TTD-based analog beamformers \(\{\tau_i\}_{i\in \left[N_{\rm{T}}\right]}\). The utility function, denoted as \(F(\cdot)\), is selected to strike a balance between overall system throughput and max-min fairness among users and is assumed to be concave, increasing and continuously differentiable. In the special case, when \(F(R_k)=R_k\), where \(R_k\) is the instantaneous downlink transmission rate of user \(k\), the objective is to maximize the sum-rate. When \(F(R_k)=\ln(R_k)\), the optimization aims at maximizing proportional fairness \cite{kelly1997charging}. The optimization problem can be formulated as follows: 
\begin{subequations}
\label{eq:max-min-1}
\begin{align}
(\mathcal{P}) 
&\max _{\left\{b_{m,k},p_{m},\tau_i, {\bm{\phi}}_{i} \right\}} \sum_{k=1}^K F (R_k) \notag\\
&\qquad\text { s.t. } 
 b_{m,k}\in \left\{0,1\right\}, \forall m\in\left[M\right] ,k\in \left[K\right], \label{max-12b}\\
&\qquad\qquad \sum_{k=1}^{K}b_{m,k} =1, \forall m\in\left[M\right], \label{max-12d} \\
&\qquad\qquad \sum_{m=1}^{M} p_m\leq P_{t}, \label{max-12e}\\
&\qquad\qquad  p_m \geq 0, \forall m\in\left[M\right], \label{max-12e1}\\
&\qquad\qquad \tau_i \in \left[0, \tau_{\max }\right],\forall i \in\left[N_{\rm{T}}\right] \label{max-12f},\\
&\qquad\qquad \left| \left[ {\bm{\phi}}_{i} \right] _j \right| =1,\forall i \!\in\!\left[N_{\rm{T}}\right], j\!\!\in\!\left[\frac{N}{N_{\rm{T}}}\right]\label{max-12g}.
% \end{split}
\end{align}
\end{subequations}
Here, \(R_{k}\), the achievable rate of user $k$, is given by
\begin{equation}
\label{eq:R1}
R_{k}=\frac{B}{M}\sum_{m=1}^{M} b_{m,k} \log _{2}\left(1+\frac{p_m\left|\mathbf{h}_{m,k}^H{\mathbf{\Phi}}\mathbf{T}_{m} \right|^{2}}{\sigma^{2}}\right).
\end{equation}
Constraint \eqref{max-12d} ensures that each subband is allocated to exactly one user. Constraint \eqref{max-12e} limits the maximum transmit power, while \eqref{max-12f} and \eqref{max-12g} address the hardware constraints of TTDs and PSs, respectively.

Note that problem \((\mathcal{P})\) is a non-convex MINLP problem due to the binary allocation constraint and highly coupled optimization variables. To deal with the problem, we propose a 3-step AO algorithm in the following section. In Section \ref{sec: DL}, the DL-based approach is proposed to achieve a computationally efficient solution for practical and real-time applications.

\section{AO Algorithm}
\label{sec: AO}
This section introduces a 3-step AO algorithm to find a near-optimal solution to the optimization problem \((\mathcal{P})\). The algorithm iteratively optimizes one of three variables: the subband allocation variable \(b_{m,k}\), the power allocation variable \(p_m\), and analog beamforming variables \(\left\{\bm{\phi}_i,\tau_i\right\}\). Additionally, the computational complexity of this approach is analyzed.

\subsection{Subband Allocation Optimization}
Initially, we specify the optimal subband assignment matrix \(b_{m,k}\) using successive convex approximation (SCA). With fixed analog beamforming vectors and power allocation variable, problem \(\mathcal{P}\) becomes a linear integer programming problem. To make it more tractable, we first convert the binary constraints into continuous constraints, enabling continuous optimization, as detailed below:
\begin{equation}
b_{m,k} \in \{0,1\} \!\iff\!\!\!\!\!\!\!\!\!\! \sum_{m\in [M] ,k\in [K]} \!\!\!\!\!\!\!\! (b_{m,k}^2 - b_{m,k}) \geq 0,~ b_{m,k} \in [0,1].
\end{equation}

Let \(\delta_{m,k}\) represent the effective channel-to-noise ratio for user \(k\) on subband \(m\), defined as \(\delta_{m,k}=\frac{\left|\mathbf{h}_{m,k}^H{\mathbf{\Phi}}\mathbf{T}_{m} \right|^{2}}{\sigma^{2}}\). The subband allocation problem is then formulated as
\begin{subequations}
\label{eq:RB}
\begin{align}
(\mathcal{P}1): &\max _{{b_{m,k},\tilde{R}_k}} ~ \sum_{k=1}^K F\left(\tilde{R}_k\right) \notag \\
&~\text { s.t. }
\tilde{R}_k\!\leq\!\frac{B}{M}\sum_{m=1}^{M} b_{m,k} \log _{2}\left(1\!+\!p_m\delta_{m,k}\right), \forall k\!\in\![K], \label{11a} \\
&~\qquad \sum_{m\in[M] ,k\in[K]} \!\!\!\!\!\!\!\left((b_{m,k})^2 - b_{m,k} \right) \geq 0, \label{constraint b}\\
&~\qquad b_{m,k}\in \left[0,1\right], \forall m\in[M],k\in[K], \label{11c}\\
&~\qquad \sum_{k=1}^{K}b_{m,k} =1, \forall m\in\left[M\right]. \label{11d}
\end{align}
\end{subequations}

The non-convexity of the above problem only lies in the constraint \eqref{constraint b}. We derive an upper bound for this constraint using the first-order Taylor expansion at point \(b_{m,k}^{(l)}\) in the \(l\)-th iteration of the SCA method:
\begin{equation}
\begin{split}
\label{eq:max-min-penalty-p-Taylor}
(b_{m,k})^2 \!-\! b_{m,k} &\geq (b_{m,k}^{(l)})^2 \!+\! 2b_{m,k}^{(l)}(b_{m,k} - b_{m,k}^{(l)}) \!-\!  b_{m,k}  \\
% &= (1 - 2b_{m,k}^{(l)})b_{m,k} + (b_{m,k}^{(l)})^2 \\
&\triangleq \Omega (b_{m,k}, b_{m,k}^{(l)}), 
\forall m\in[M] ,k\in [K].
\end{split}
\end{equation}

By replacing the non-convex constraint with \eqref{eq:max-min-penalty-p-Taylor} and incorporating it into the objective function as a penalty term, problem \(\mathcal{P}1\) can be reformulated as 
\begin{equation}
\label{eq:RB1}
\begin{split}
(\mathcal{P}1.1): ~ 
\max _{{b_{m,k},\tilde{R}_k}} &~ \sum_{k=1}^K F\left(\tilde{R}_k\right) + \rho \Omega (b_{m,k}, b_{m,k}^{(l)}),  \\
\text { s.t. } &~
\eqref{11a}, \eqref{11c}, \eqref{11d} .\notag %, \eqref{max-12f}, \eqref{max-12e1}  ,
\end{split}
\end{equation}
% \eqref{eq:RB1}
with $\rho > 0$ being the penalty factor. Here, $\rho$ is initialized with a small value to ensure significant emphasis on maximizing the total concave utility function associated with user rates and then is gradually increased to increase the influence of the penalty term. Problem \((\mathcal{P}1.1)\) is convex whose stationary-point solution can be efficiently obtained by convex optimization toolboxes \cite{grant2014cvx}.

\subsection{Analog Beamforming Optimization}
Given a fixed subband allocation and power distribution, we formulate the optimization problem for analog beamforming matrices as follows:
\begin{equation}
\label{eq:analog}
\begin{split}
&(\mathcal{P}2): \\
 &\max _{\bm{\phi}_i,\tau_i} ~\sum_{k=1}^K F\left(\frac{B}{M}\sum_{m=1}^{M} b_{m,k} \log _{2}\left(1\!+\!\frac{p_m\left|\mathbf{h}_{m,k}^H{\mathbf{\Phi}}\mathbf{T}_{m} \right|^{2}}{\sigma^{2}}\right)\right), \\
&~~\text { s.t. } ~  \eqref{max-12f}, \eqref{max-12g}.
\end{split}
\end{equation}

Because of the complex coupling between PS-based and TTD-based variables, problem \((\mathcal{P}2)\) becomes challenging to solve. To overcome this, we introduce an auxiliary variable $\mathbf{w}_m=\mathbf{\Phi} \mathbf{T}_{m}$ representing the optimal analog beamformer. This optimal analog beamformer is determined by solving
\begin{equation}
\label{eq:analog1}
\begin{split}
(\mathcal{P}2.1):& \\
\max _{\mathbf{w}_m} & ~\sum_{k=1}^K F\left(\frac{B}{M}\sum_{m=1}^{M} b_{m,k} \log _{2}\left(1\!+\!\frac{p_m\left|\mathbf{h}_{m,k}^H\mathbf{w}_m \right|^{2}}{\sigma^{2}}\right)\right), \\
\text { s.t. } ~ & \left|\left[\mathbf{w}_m\right]_n\right|=\frac{1}{\sqrt{N}}, \forall m\in[M],n\in[N],
\end{split}
\end{equation}
where \(\mathbf{w}_m\) can be directly optimized as
\begin{equation}
\label{optimal w}
\mathbf{w}_m = \frac{1}{\sqrt{N}}\sum_{k=1}^Kb_{m,k}\exp\left(j\angle{\mathbf{h}_{m,k}}\right),\forall m\in[M].
\end{equation}

Subsequently, the PS-based and TTD-based analog beamformers, \(\bm{\phi}_i\) and \(\tau_i\), are optimized to approximate \(\mathbf{w}_m\). The optimization problem yields
\begin{equation}
\begin{split}
(\mathcal{P}2.2): ~ \min_{\bm{\phi}_i,\tau_i} ~ &\left\| \mathbf{w}_m - \mathbf{\Phi}\mathbf{T}_{m} \right\|^{2} ,\\
\text { s.t. } ~ & \eqref{max-12f}, \eqref{max-12g},
\end{split}
\end{equation}
which can then be solved with block coordinate descent algorithm \cite{ratnam2022joint}. Decomposing this problem, each PS-based analog beamformer \(\bm{\phi}_i\) can be independently optimized by
\begin{equation}
\label{eq:max-min-PS}
\begin{split}
(\mathcal{P}2.2.1): ~ \max _{{\bm{\phi}}_i} & \sum_{i=1}^{N_{\rm{T}}} \sum_{m=1}^{M} \mathrm{Re}\left\{ \widetilde{\mathbf{w}}_{m,i}^H \bm{\phi}_{i}e^{-j 2 \pi f_{m} \tau_{i}} \right\},\\
\text { s.t. } & \left| \left[ {\bm{\phi}}_{i} \right] _j \right| =1,\forall i \!\in\!\left[N_{\rm{T}}\right], j\!\!\in\!\left[\frac{N}{N_{\rm{T}}}\right],
\end{split}
\end{equation}
where \(\widetilde{\mathbf{w}}_{m,i}=\left[\mathbf{w}_m\right]_{(i-1)\frac{N}{N_{\rm{T}}}+1 : i\frac{N}{N_{\rm{T}}} }, \forall i \in \left[N_{\rm{T}}\right]\). The optimal solution for \((\mathcal{P}2.2.1)\) with fixed time delay \(\tau\) is obtained as
\begin{equation}
\label{optimal ps}
{\bm{\phi}}_i = e^{j\angle \left(\sum_{m=1}^{M}\widetilde{\mathbf{w}}_{m,i}e^{j2\pi f_{m}\tau_{i}}\right) } ,\forall i \!\in\!\left[N_{\rm{T}}\right].
\end{equation}

Similarly, the TTD-based analog beamforming optimization becomes
\begin{equation}
\label{eq:max-min-penalty-TTD}
\begin{split}
(\mathcal{P}2.2.2): ~ \max _{\tau_{i}} & \sum_{i=1}^{N_{\rm{T}}} \sum_{m=1}^{M} \mathrm{Re}\left\{ \widetilde{\mathbf{w}}_{m,i}^H{\bm{\phi}}_{i}e^{-j 2 \pi f_{m} \tau_{i}} \right\},\\
\text { s.t. } & \tau_i \in \left[0, \tau_{\max }\right],\forall i \in\left\{1, \ldots, N_{\rm{T}}\right\}.
\end{split}
\end{equation}
This non-convex problem is a single-variable optimization within a fixed interval and can be efficiently solved via the linear search approach within the finite interval \([0,\tau_{\max}]\). Specifically, denote the search step as \(I_{\rm{T}}\), the search set can be given as \(\mathcal{T} = \left\{ 
0,\frac{\tau_{\max}}{I_{\rm{T}}-1},\ldots ,\frac{(I_{\rm{T}}-2)\tau_{\max}}{I_{\rm{T}}-1}, \tau_{\max}\right\}\). The near-optimal solution is then obtained as 
\begin{equation}
\label{eq:solution-TTD}
\tau_i = \text{arg}\!\max_{\tau_i\in\mathcal{T}}\sum_{m=1}^{M} \mathrm{Re}\left\{ \widetilde{\mathbf{w}}_{m,i}^H{\bm{\phi}}_{i}e^{-j 2 \pi f_{m} \tau_{i}} \right\} ,\forall i \!\in\!\left[N_{\rm{T}}\right].
\end{equation}

\subsection{Power Allocation Optimization}
With given subband allocation and analog beamformers, the optimization problem for power allocation is formulated as
\begin{equation}
\label{eq:power}
\begin{split}
(\mathcal{P}3): ~
\max _{p_m} ~ &\sum_{k=1}^K F\left(\frac{B}{M}\sum_{m=1}^{M} b_{m,k} \log _{2}\left(1\!+\!p_m\delta_{m,k}\right)\right), \\
\text { s.t. } ~ & \eqref{max-12e}, \eqref{max-12e1}.
\end{split}
\end{equation}
which is a convex optimization problem. The Lagrangian associated with the problem is
\begin{equation}
\label{eq:power Lagrangian}
\begin{split}
\mathcal{L}(p_m)=&-\sum_{k=1}^K F\left(\frac{B}{M}\sum_{m=1}^{M} b_{m,k} \log _{2}\left(1\!+\!p_m\delta_{m,k}\right)\right) \\
&- \sum_{m=1}^{M}\lambda_m p_m + \lambda_0 \left(\sum_{m=1}^{M} p_m - {P_t}\right),
\end{split}
\end{equation}
where \(\left\{\lambda_i\right\}_{i\in \{0,\ldots,M\}}\) are the dual variables. By differentiating the Lagrangian in \eqref{eq:power Lagrangian} with respect to \(p_m\) and substituting the result into the Karush-Kuhn-Tucker conditions \cite{boyd2004convex}, the optimal power allocation can be derived following the standard water-filling approach \cite{goldsmith2005wireless}. Specifically, under a linear utility function \(F(R_k)=R_k\), the optimal solution is
\begin{equation}
p_m = \sum_{k=1}^K b_{m,k}\left( \frac{B}{\lambda_0M\ln2}-\frac{1}{\delta_{m,k}} \right)^+.
\end{equation}
Here, \(\left(x\right)^+ \triangleq \max\left(x,0\right)\), and \(\frac{\lambda_0M\ln2}{B}\) is the water level cut-off that satisfies
\begin{equation}
\sum_{k=1}^K\sum_{m=1}^M b_{m,k}\left( \frac{B}{\lambda_0M\ln2}-\frac{1}{\delta_{m,k}} \right)^+ = {P_t}.
\end{equation}

For a logarithmic utility function \(F(R_k)=\ln(R_k)\), the power allocation variables can be obtained using subgradient methods or barrier techniques combined with Newton's method for optimization \cite{boyd2004convex}.

% the optimal power allocation satisfies
% \begin{equation}
% \label{eq:power Lagrangian solution}
% \left[-\sum_{k=1}^K \frac{\frac{1}{\ln2}\cdot\frac{b_{m,k}\delta_{m,k}}{1+p_m\delta_{m,k}}}{\sum_{m=1}^{M} b_{m,k} \log _{2}\left(1\!+\!p_m\delta_{m,k}\right)} + \lambda_0\right] p_m=0.
% \end{equation}
% This coupling of power allocation variables can be resolved using subgradient methods or barrier techniques combined with Newton's method for optimization. \cite{boyd2004convex}.

\subsection{Overall Algorithm and Complexity Analysis}
The comprehensive algorithm for addressing problem \eqref{eq:max-min-1} is detailed in \textbf{Algorithm \ref{alg}}. The computational complexity of subband allocation optimization within \textbf{Algorithm \ref{alg}} is predominantly governed by the SCA process, estimated as \(\mathcal{O}\left(I_{\rm{SCA}} M^3K^3 \right)\), where \(I_{\rm{SCA}}\) represents the number of SCA iterations. The complexities for updating \(\bm{\phi}_i\) and \(\tau_i\) are \(\mathcal{O}\left(MN \right)\) and \(\mathcal{O}\left(MN_{\rm{T}}I_{\rm{T}} \right)\), respectively. For power distribution optimization, the complexity varies by the utility function: \(\mathcal{O}\left(M\log(M) \right)\) under a linear utility function \(F(R_k)=R_k\), and \(\mathcal{O}\left(M\log(1+\frac{1}{\epsilon_{\rm{NT}}}) \right)\) for logarithmic utility function \(F(R_k)=\ln(R_k)\), where \(\epsilon_{\rm{NT}}>0\) specifies the precision of Newton’s method. Assuming the maximum number of iterations for AO and updating analog beamformers as \(I_{\rm{AO}}\) and \(I_{\rm{AN}}\) respectively, the total computational complexity of \textbf{Algorithm \ref{alg}} is thus \(\mathcal{O}(I_{\rm{AO}} (I_{\rm{SCA}} M^3K^3 + I_{\rm{AN}}(MN + MN_{\rm{T}}I_{\rm{T}}) + M\log(1+\frac{1}{\epsilon_{\rm{NT}}}) ) )\).

\begin{algorithm}[htbp]
    \caption{AO algorithm for solving problem \eqref{eq:max-min-1}.}
    \label{alg}
    \renewcommand{\algorithmicrequire}{\textbf{Input: }}
    \renewcommand{\algorithmicensure}{\textbf{Output:}}
    \begin{algorithmic}[1]
        \Require $\mathbf{h}_{m,k}$, $K_{\mathrm{NU}}$, $K_{\mathrm{FU}}$, $N_{\rm{T}}$, $\tau_{\max }$, $\rho$, ${\epsilon}$, \(l_{\rm{max}}\).
        %%input
        \Ensure The power allocation variable \(p_m\), the subband allocation variable \(b_{m,k}\), the PS-based beamformer \(\bm{\phi}_i\), and the TTD-based beamformer \(\tau_i\).
        %%output
        \State  Initialize $\rho=10^{-5}$, $\bm{\phi}_i$, $\tau_i$, and \(p_m = \frac{P_{\rm{t}}}{M}\).
        \Repeat
        \State  Update \(b_{m,k}\) through SCA.
        \State  Update \(\mathbf{w}_{m}\) with \eqref{optimal w}.
            \Repeat
            \State Update ${\bm{\phi}}_i$ with \eqref{optimal ps}.
            \State Update $\tau_i$ with \eqref{eq:solution-TTD}.
            \Until the fractional decrease of the objective function value is below a predefined threshold \(\epsilon\) or the number of iterations reaches the maximum value \(l_{\rm{max}}\).
        \State Update $p_m$ with water-filling.
        \State Update $\rho = 5\rho$.
        \Until the constraint violation of the penalty term is below the predefined threshold \({\epsilon}\) or the number of iterations reaches the maximum value \(l_{\rm{max}}\).
        % \RETURN EEEEE
    \end{algorithmic}
\end{algorithm}
 
\section{GAT-Enabled Beamforming Design}
\label{sec: DL}
\begin{figure*}[htbp]
    \centering    
    \includegraphics[width=1\linewidth]{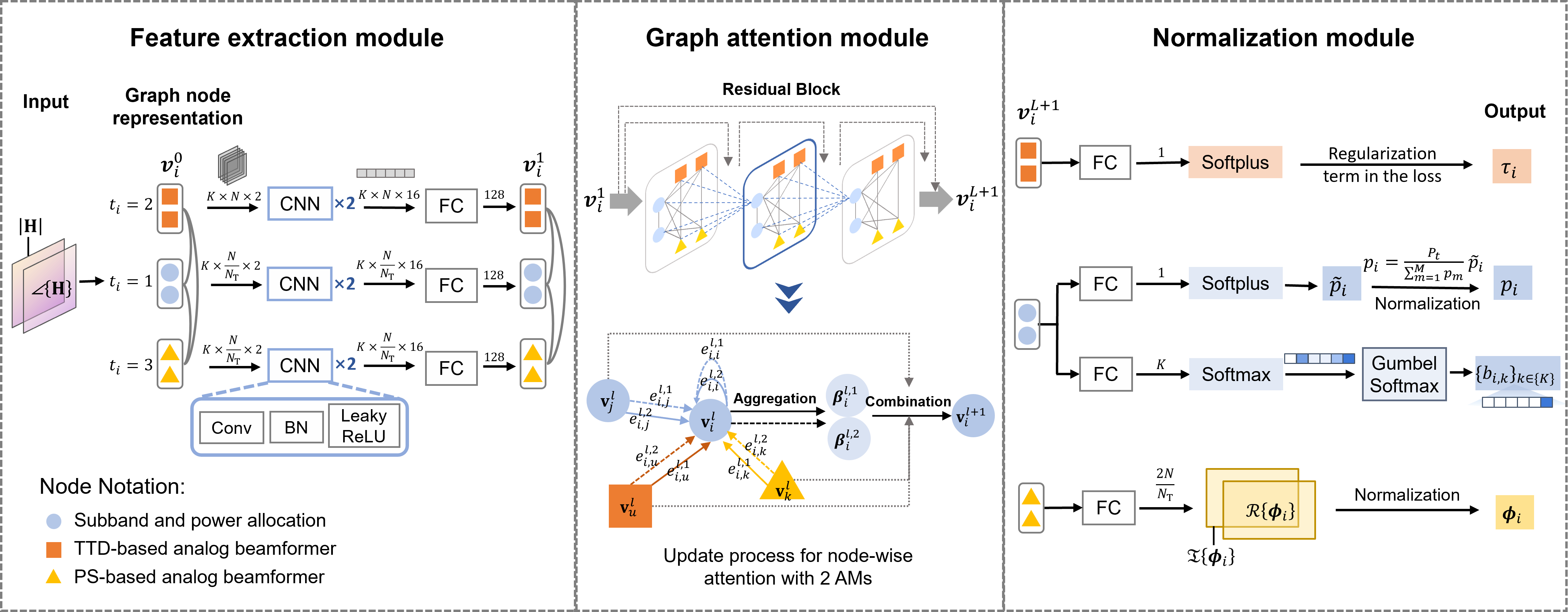}
    \caption{Structure of the proposed network architecture, comprising a 2-layer CNN for feature extraction, a 3-layer graph attention module and a normalization module for subband allocation and beamforming optimization.}
    \label{fig: GAT}
\end{figure*}
In this section, we propose a GAT-based learning approach, which is low in complexity and feasible for real-time implementation. As depicted in Fig. \ref{fig: GAT}, the overall network is composed of a feature extraction module, a graph attention module, and a normalization module. The input to the network is a real-valued tensor of the channel, with the amplitude and angle parts stored separately. The output are the power allocation variable \(p_m\), the subband allocation variable \(b_{m,k}\), the PS-based beamformer \(\bm{\phi}_i\), and the TTD-based beamformer \(\tau_i\).

\begin{table*}[htpb]
\renewcommand\arraystretch{1.1}
\centering
\caption{The definition, input, and output of nodes in GAT. }
\label{GATnode}
\begin{tabular}{|c|c|c|c|c|c|}
\hline
\textbf{Node} & \textbf{Node Index} & \textbf{Type} & \textbf{Definition} & \textbf{Input} & \textbf{Normalized Output}\\
\hline
\multirow{4}{*}{\(\mathbf{v}_i\)} & \(1\leq i \leq M\) & 1 & Subband and power allocation & \(\left|{\mathbf{H}_i}\right|\oplus_3 \angle{\mathbf{H}_i}\in \mathbb{R}^{K \times N\times 2}\) & \(p_i,\{b_{i,k}\}_{k\in \left[K\right]}\) \\
\cline{2-6} 
 & \(M\!+\!1\leq i \leq M\!+\!N_{\rm{T}}\) & 2 & PS-based beamformer & \(\left|\mathbf{H}_{i}^{\rm{sub}}\right|\oplus_3 \angle{\mathbf{H}_{i}^{\rm{sub}}}\in \mathbb{R}^{K \times \frac{N}{N_{\rm{T}}}\times 2}\) & \({\bm{\phi}}_i\)\\
\cline{2-6}
 & \(M\!+\!N_{\rm{T}}\!+\!1 \!\leq\! i \!\leq M\!+\!2N_{\rm{T}}\) & 3 & TTD-based beamformer & \(\left|\mathbf{H}_{i}^{\rm{sub}}\right|\oplus_3 \angle{\mathbf{H}_{i}^{\rm{sub}}}\in \mathbb{R}^{K \times \frac{N}{N_{\rm{T}}}\times 2}\)& \(\tau_i\)\\
\hline
\end{tabular}
\end{table*}
\subsection{Graphical Representation}

In the JPTA OFDM system, each subband is equipped with a unique resource allocation, while TTD-based and PS-based analog beamformers are shared across all subbands. Unlike FC networks, the GAT captures the interactions between resource allocation and analog beamformers by leveraging the established topology of the antenna architecture. This approach ensures permutation invariance and permutation equivariance within the optimization problem. Permutation invariance guarantees that the configuration of the analog beamformers is independent of the ordering of the subbands, while permutation equivariance ensures that any permutation of the subbands is correspondingly reflected in the resource allocation matrices. Additionally, unlike conventional GNNs that apply uniform weights to all nodes, the node-wise GAT assigns distinct weights to different nodes, enhancing its ability to accurately model the interactions between various nodes.

As shown in Fig. \ref{fig: GAT}, the considered JPTA can be modeled as an undirected graph \( \mathcal{G} = (\mathcal{V}, \mathcal{E}) \) with a set of nodes \( \mathcal{V}\) and a set of edges \( \mathcal{E}\). The nodes are classified into three types, corresponding to different components of the system: subband and power allocation \(\left\{\mathbf{v}_i\right\}_{i\in\{1,\ldots,M\}}\), PS-based analog beamformers \(\left\{\mathbf{v}_i\right\}_{i\in\{M+1,\ldots,M+N_{\rm{T}}\}}\), and TTD-based analog beamformers \(\left\{\mathbf{v}_i\right\}_{i\in\{M+N_{\rm{T}}+1,\ldots,M+2N_{\rm{T}}\}}\). The channel across different users on subband \(m\) is denoted by \(\mathbf{H}_m=\left[ \mathbf{h}_{m,1}^H; \ldots; \mathbf{h}_{m,K}^H \right]\in \mathbb{C}^{K \times N \times 1}\). For each TTD-connected sub-array antenna, the channel matrix for subband \(m\) on TTD line \(i\) can be represented by \(\left[\mathbf{H}_m\right]_{:,\frac{(i-1)N}{N_{\rm{T}}}:\frac{iN}{N_{\rm{T}}},:}\), where \(i \in \left[N_{\rm{T}}\right]\). The input for the beamformers is calculated by averaging across different subbands, expressed as
\begin{equation}
\mathbf{H}_{i}^{\rm{sub}}=\sum_{m=1}^M \left[\mathbf{H}_m\right]_{:,\frac{(i-1)N}{N_{\rm{T}}}:\frac{iN}{N_{\rm{T}}},:}, i \in \left[N_{\rm{T}}\right].
\end{equation}

The node type, definition, inputs, and normalized outputs of each node within the DL network are detailed in Table \ref{GATnode}. The interactions between node \(i\) and node \(j\) within this architecture are captured by the learnable adjacency matrix \(\mathcal{A}_{i,j}\), which is initialized as
\begin{equation}
\mathcal{A}_{i,j}=
\begin{cases}
10^{-2}, &i\neq j,M< i,j \leq M\!+\!2N_{\rm{T}};\\
f_i/f_c,&1\!\leq i \leq \!M,M\!+\!N_{\rm{T}}< j \leq \!M\!+\!2N_{\rm{T}};\\
1, &\text{otherwise},
\end{cases}
\end{equation}
This configuration allows for the translation of time delays into phase adjustments across different subbands by multiplying TTD node parameters with corresponding frequency coefficients, enhancing the network's capability to learn the complex interactions between TTD nodes and resource allocation nodes. Notably, the adjacency matrix is symmetric and shared across all layers, i.e., \(\mathcal{A}_{i,j} = \mathcal{A}_{j,i}\).

\subsection{Feature Extraction Module}
Two CNN blocks followed by an FC layer are employed to extract features from the channel matrices. Each CNN block consists of a convolution (Conv) layer with a kernel size of 3 and a stride of 1, followed by a batch normalization (BN) layer and a Leaky ReLU (LReLU) activation function. The LReLU function introduces small and non-zero gradients when the unit is inactive, formulated as:
\begin{equation}
\text{LReLU}(x)=
\begin{cases}
x, ~ &x>0;\\
\alpha_{\rm{LR}}x,~ &x\leq 0,
\end{cases}
\end{equation}
where \(\alpha_{\rm{LR}}\) denotes the negative slope and is set to 0.1 in the training process. The operation of each CNN block can be expressed as
\begin{equation}
\text{CNN}(\mathbf{X})=\text{LReLU}\left(\text{BN}\left(\mathbf{W}_{\rm{CNN}}*\mathbf{X}+\mathbf{B}_{\rm{CNN}}\right)\right),
\end{equation}
where \(\mathbf{X}\) is the real-valued input tensor, \(*\) denotes the convolution operation, \(\mathbf{W}_{\rm{CNN}}\) and \(\mathbf{B}_{\rm{CNN}}\) represent the weight and bias of the convolution layer. 

The three-dimensional output tensor from the CNN blocks is then vectorized into a one-dimensional tensor and fed into the FC layer, which can be represented by 
\begin{equation}
\text{FC}(\mathbf{X})=F_{\rm{ac}}\left(\mathbf{W}_{\rm{FC }}\mathbf{X}+\mathbf{b}_{\rm{FC}}\right),
\end{equation}
where \(\mathbf{W}_{\rm{FC}}\) and \(\mathbf{b}_{\rm{FC}}\) are the weight and bias variable, \(F_{\rm{ac}}\) denotes the activation function.

\subsection{Node-Wise Graph Attention Module}
Different types of nodes in a graph play distinct roles and exhibit varying levels of importance in learning node embeddings for the beamforming task. To this end, we implement a node-wise attention mechanism (AM), which assesses the significance of each neighboring node and aggregates the representations of these relevant neighbors to form a node embedding. The node-wise graph attention module consists of \(L\) graph attention layers, each of which comprises an aggregation and combination process, as illustrated in Fig. \ref{fig: GAT}.
\subsubsection{Aggregation} During the aggregation phase, an AM is adopted to enhance the learning capability. At the \(l\)-th attention layer, input features \(\mathbf{v}_{i}^{l} \in \mathbb{R}^{F^{l}}\) are processed and \(X(l)\) AMs are applied to compute the attention coefficients. The attention coefficients indicating the importance of node \(j\)'s feature to node \(i\) after the \(x\)-th AM at layer \(l\) is given by 
\begin{equation}
\begin{split}
e_{i,j}^{l,x}=&\mathcal{A}_{i,j}\text{LReLU}\left(\mathbf{Z}_{t_i}^{l,x}\left(\mathbf{W}_{t_i}^{l,x}\mathbf{v}_{i}^l \oplus_2 \mathbf{W}_{t_j}^{l,x}\mathbf{v}_{j}^l \right)\right), \\
&l\in \left[L\right], x\in \left[X(l)\right],
\end{split}
\end{equation}
where \(t_i\) denotes the node type of node \(i\). \(\mathbf{Z}_{t_i}^{l,x}\in \mathbb{R}^{1 \times 2F^{l+1}}\), \(\mathbf{W}_{t_i}^{l,x} \in \mathbb{R}^{F^{l+1} \times F^l}\), and \(\mathbf{W}_{t_j}^{l,x} \in \mathbb{R}^{F^{l+1} \times F^l}\) are type-specific weight matrices. 

Then the aggregated feature for node \(i\) is given by
\begin{equation}
\bm{\beta}_{i}^{l,x}=\sum_{j \in [V_{i}] }\frac{\exp{(e_{i,j}^{l,x})}}{\sum_{k \in [V_{i}]} \exp{(e_{i,k}^{l,x})}} \mathbf{W}_{t_j}^{l,x} \mathbf{v}_{j}^{l}.
\end{equation}
where \([V_{i}]\) denotes the set of neighboring nodes of node \(i\).
% Attention scores \(\alpha_{i,j}^{l,x}\) are normalized across the nodes' neighbors using a softmax function:
% \begin{equation}
% \alpha_{i,j}^{l,x}=\frac{\exp{(e_{i,j}^{l,x})}}{\sum_{k \in [V_{i}]} \exp{(e_{i,k}^{l,x})}},
% \end{equation}
% where \([V_{i}]\) denotes the set of neighboring nodes (including \(i\))  of node \(i\). The aggregated feature for node \(i\) is then given by
% %\begin{small}
% \begin{equation}
% \bm{\beta}_{i}^{l,x}=\sum_{j \in [V_{i}] }\alpha_{i,j}^{l,x} \mathbf{W}_{t_j}^{l,x} \mathbf{v}_{j}^{l}.
% \end{equation}

\subsubsection{Combination}
The node features are updated by combing the \(X(l)\) aggregated features from each AM and the input node features of the preceding graph attention layer:
\begin{equation}
\mathbf{v}_{i}^{l+1} \!\!=\!\! 
\begin{cases}
F_{\rm{ac}}\!\left(F_{\rm{concat}}\! \left( \left\{\bm{\beta}_{i}^{l,x}\right\}_ {x \in X(l) }\right) \!\!+\!\! \mathbf{v}_{i}^{l}\Theta^l \right), \!\!\!\!&l \!\in\! [L-1],\\
F_{\rm{ac}}\left(\sum_{x \in X(l) }\bm{\beta}_{i}^{l} \!+\! \mathbf{v}_{i}^{l}\Bar{\Theta}^l \!+\!\mathbf{v}_{i}^{1}\tilde{\Theta}^l\right), &l=L,
\end{cases}
\end{equation}
where \(\Theta^l \in \mathbb{R}^{F^l \times X(l)F^{l+1}}\), \(\Bar{\Theta}^l \in \mathbb{R}^{F^l \times F^{l+1}}\), and \(\tilde{\Theta}^l \in \mathbb{R}^{F^1 \times F^{l+1}}\) denote the trainable parameters of the feedforward network for the layer-wise residual and network-wise residual. \(F_{\rm{concat}}\) represents the concatenation function.

\subsection{Normalization Module}
This module employs multiple FC layers and normalization layers to transform the output features \(\left\{\mathbf{v}_{i}^{L+1}\right\}_{i\in [M+2N_{\rm{T}}]}\) to the required allocation and beamforming vectors. The normalization methods for different nodes are outlined as follows:

\subsubsection{Subband Allocation Variable} To address the binary constraint specified in \eqref{max-12b}, the Gumbel-Max trick is employed to facilitate a differentiable approximation \cite{jang2017categorical}. Initially, an FC layer with an output length of \(K\) and a softmax activation function maps \(\{\mathbf{v}_{i}^{L+1}\}_{i\in[M]}\) into a probability distribution vector $\mathbf{q}_i = [q_{i,1}, \ldots, q_{i,K}]^T$ for $ i \in[M]$, representing the likelihood of allocating subband $i$ to each user. Subsequently, the Gumbel-softmax function is applied to $q_{i,k}$, resulting in 
\begin{equation}
G(q_{i,k}) = \frac{\exp\left(\left(\log(q_{i,k}) + g_{i,k}\right)/\mu\right)}{\sum_{j=1}^K \exp\left(\left(\log(q_{i,j}) + g_{i,j}\right)/\mu\right)},i\!\in\![M], k\!\in\![K],
\end{equation}
where $g_{i,k}$ is sampled from the Gumbel distribution and $\mu$ denotes the softmax temperature. As $\mu \to 0$, $G(\mathbf{q}_i)$ converges to a categorical distribution, effectively mirroring the discrete nature of the subband allocation variable.

\subsubsection{Power Allocation Variable and Analog Beamformers} 
As depicted in Fig. \ref{fig: GAT}, a normalization layer is employed to scale the outputs, ensuring compliance with the power constraints specified in \eqref{max-12e}, the time delay constraints in \eqref{max-12f}, and the phase constraints in \eqref{max-12g}. Specifically, the power allocation matrices are normalized as 
\begin{equation}
p_i= \frac{{P_t}}{{\sum_{i=1}^M\tilde{p}_i}}\tilde{p}_i, \forall i \in \{1,\ldots,M\},
\end{equation}
where \(\tilde{p}_i=\text{softplus}(\text{FC}\left(\mathbf{v}_{i}^{L+1})\right)\). The softplus function ensures the outputs are always positive, i.e., \(\tilde{p}_i\geq0\), which can be formulated as
\begin{equation}
\text{softplus}(x) = \ln (1+e^x).
\end{equation}
% the output from a FC layer. 

For PS-based analog beamformers, the phase shifts are derived by combining the real and imaginary components of the output from a FC layer and normalizing them. The mathematical representation of this process is as follows:
\begin{equation}
\begin{split}
&\text{vec}\left(\left[\tilde{\bm{\phi}}_i^{\mathrm{Re}},\tilde{\bm{\phi}}_i^{\mathrm{Im}}\right]\right)=\text{FC}(\mathbf{v}_{i}^{L+1}), \\
&{\bm{\phi}}_i=\frac{\tilde{\bm{\phi}}_i^{\mathrm{Re}}+j\tilde{\bm{\phi}}_i^{\mathrm{Im}}}{\|\tilde{\bm{\phi}}_i^{\mathrm{Re}}+j\tilde{\bm{\phi}}_i^{\mathrm{Im}}\|}\in \mathbb{C}^{\frac{N}{N_{\rm{T}}} \times 1},
\\
&~~~~~ \forall i \!\in\! \{M\!+\!1,\! \ldots,\! M\!+\!N_{\rm{T}}\}.
\end{split}
\end{equation}

For the TTD-based analog beamformers, the time delays are adjusted using a softplus activation function and reluarized in the loss, ensuring that the resulting time delays do not exceed the maximum allowable values \(\tau_{\max}\).

\subsection{Loss Function and Complexity Analysis}
The proposed network is trained in an unsupervised manner, where the loss function design plays a key role. Our proposed loss function incorporates the optimization target and regularization terms for time delays of TTDs and the one-hot vector, which can be expressed as
\begin{equation}
\begin{split}
L\! = \!& -\! \sum_{k=1}^K\! F(R_k) \!+\! \lambda_1\!\!\sum_{i=1}^{N_{\rm{T}}}\psi(\tau_i) \!+\!  \lambda_2\! \sum_{i\neq j}\left|\sum_{k=1}^K b_{i,k} b_{j,k}\right|^2 \\
& - \lambda_3\sum_{m=1}^M \sum_{k=1}^K b_{m,k} \log_2 b_{m,k}, 
\end{split}
\end{equation}
where \(\psi(x) = \max\{0,x-\tau_{\max}\}\) is the regularization term to force the time delay of TTD not to exceed the maximum allowable values. Parameters \(\lambda_1\), \(\lambda_2\), and \(\lambda_3\) are weight factors and remain fixed during the training phase. The third term pushes the subband allocation variable \(b_{m,k}\) satisfying constraint \eqref{max-12d}, while the fourth term drives the values of \(\{b_{m,k}\}_{m \in \left[M\right],k \in \left[K\right]}\) closer to 0 or 1.

Since the network training can be conducted off-line, we only consider the computational complexity of the inference stage. The computational complexity of the proposed GAT-based network primarily stems from the CNN-based feature extraction module and the graph attention module. Specifically, the complexity of the CNN layers can be approximated by \(2KNN_{\rm{K}}^2\Bar{N}_{\rm{C}}^2\), where \(N_{\rm{K}}\) represents the kernel size and \(\Bar{N}_{\rm{C}}\) denotes the average number of channels per layer. For the graph attention module, the complexity can be estimated as \(\sum_{l=1}^L\left(3X(l)F^{l+1}(2+F^l)\right)+\sum_{l=1}^{L-1}X(l)F^lF^{l+1}+(F^1+F^L)F^{L+1}+{(1+M+2N_{\rm{T}})(M+2N_{\rm{T}})}/{2}\).

\section{Numerical Results}
\label{sec: results}
\begin{table}[t]
\centering
\caption{Main notations and their typical values}
\label{para}
\begin{tabular}{|c| c| c|}
\hline
\textbf{Notation} &\textbf{Definition} & \textbf{Value} \\
\hline
$f_c$ &  { Carrier frequency } & 100 GHz \\
\hline
$B$ &  { System bandwidth } & 10 GHz \\
\hline
$N$ &  { Number of antennas at BS} & 64 \\
\hline
$P_t$ &  { Transmit power at BS} & 40 dBm \\
\hline
$\sigma^2$ &  {Noise power density} & -174 dBm/Hz \\
\hline
$M$ &  { Number of subbands} & 16 \\
\hline
$N_{\rm{T}}$ &  { Number of TTDs} & 16 \\
\hline
$\tau_{\rm{max}}$ &  { Maximum time delay of TTDs} & 5 ns  \\
\hline
$I_{\rm{T}}$ &  { Linear search step for TTDs} & 2000  \\
\hline
$l_{\rm{max}}$ &  {Mximum number of iterations of AO} & \(30\)  \\
\hline
$\epsilon$ &  { Optimization threshold of AO} & \(10^{-5}\)  \\
\hline
% $P_{\rm{T}}$ &  { Power of one TTD} & 100mW  \\
% \hline
% $P_{\rm{PS}}$ &  { Power of one PS} & 30mW  \\
% \hline
% $P_{\rm{RF}}$ &  { Power of one RF chain} & 200mW  \\
% \hline
$\lambda_1, \lambda_2, \lambda_3$ &  {Penalty parameter in loss function} & \(10^{10}\), 0.5, 1  \\
\hline
\end{tabular}
\end{table}
\begin{table}[t]
\centering
\caption{Structure of proposed GAT}
\label{network}
\begin{tabular}{|c| c| c| c| c| c|}
\hline
\textbf{Layer} &\textbf{Input} &\textbf{Output} &\textbf{AMs} &\textbf{LReLU} &\textbf{Residual}\\
\hline
1 & 128 & 64 & 4 & \checkmark & \checkmark \\
\hline
2 & 64*4 & 128 & 4 & \checkmark & \checkmark \\
\hline
3 & 128*4 & 256 & 2 & \checkmark & \checkmark \\
\hline
\end{tabular}
\end{table}
In this section, we present the numerical results of the proposed JPTA-based beamforming schemes. Unless otherwise specified, the simulations adhere to the parameters listed in Table \ref{para}. Notably, the Rayleigh distance for our simulation setup is \(5.96\) meters. We assume that users are randomly distributed within a half-circular area, spanning from 0 to 180 degrees, sampled at intervals of 0.5 degrees around the BS. The users are located at distances ranging from  \(1\) to \(20\) meters from the BS, with increments of \(0.5\) meters. For model training, we generated 10,000 samples as the training set, 2,000 as the validation set, and 50 as the test set in both 2-user and 5-user scenarios.

The detailed architecture of the proposed GAT is shown in Table \ref{network}. The implementation was conducted using Pytorch. Throughout the training phase, we utilized the Adam optimizer over 1000 epochs with a progressively increasing batch size of \(\{8,32,128\}\) and a learning rate of \(10^{-4}\).

For comparison, we consider the following two baselines:
\begin{itemize}
    \item \textbf{FD beamforming with subband allocation}: This is a FD beamforming design, where each antenna is connected to a dedicated RF chain and it establishes a performance upper bound for all analog beamforming schemes. The near optimal solution of the joint FD beamforming and subband allocation is found using iterative SCA and interior point optimization techniques.  
    \item \textbf{PA beamforming with subband allocation}: This benchmark represents the conventional PS-only beamforming architecture without TTD components. This PA architecture is equivalent to the JPTA with \(N_{\rm{T}}=0\). The subband allocation matrices and analog beamformers for this architecture are optimized through the proposed AO algorithm and DL network. 
\end{itemize}

\subsection{Comparison Between AO and GAT Methods}
Before we compare the proposed JPTA beamforming methods with baseline approaches, we illustrate in this subsection the internal comparison between the optimization-based AO algorithm in Section \ref{sec: AO} and the learning-based GAT method in Section \ref{sec: DL}.

\begin{table*}[htbp]
\renewcommand\arraystretch{1.3}
\centering
\caption{Complexity and logarithmic rate performance comparison}
\label{Table complexity}
\begin{tabular}{|c| c| c| c| c|}
\hline
\multirow{2}{*}{\textbf{Technique}} & \multirow{2}{*}{\textbf{Computational complexity}}  & {\textbf{Average CPU}} &\textbf{Logarithmic} &\textbf{Logarithmic rate for} \\
& & \textbf{run time} &\textbf{rate for PA} &\textbf{JPTA when \(N_{\rm{T}}=16\)} \\
\hline
AO & \(\mathcal{O}(I_{\rm{AO}} (I_{\rm{SCA}} M^3K^3 + I_{\rm{AN}}(MN + MN_{\rm{T}}I_{\rm{T}}) + M\log(1+\frac{1}{\epsilon_{\rm{NT}}}) ) )\) & 7.11 min & 118.382 &118.531  \\
\hline
\multirow{2}{*}{DL} & \(2KNN_{\rm{K}}^2\Bar{N}_{\rm{C}}^2\sum_{l=1}^L\left(3X(l)F^{l+1}(2+F^l)\right)+\sum_{l=1}^{L-1}X(l)F^lF^{l+1}\)&\multirow{2}{*}{0.11 s} &\multirow{2}{*}{118.380} &\multirow{2}{*}{118.525}\\
&\(+(F^1+F^L)F^{L+1}+{(1+M+2N_{\rm{T}})(M+2N_{\rm{T}})}/{2}\) & && \\
\hline
\end{tabular}
\end{table*}
Table \ref{Table complexity} summarizes the approximate complexities and logarithmic rate performance of AO and GAT. Both methods were run on the same Intel Xeon Gold 6342 CPU @ 2.80 GHz for a fair comparison. As the running time in the offline training stage is normally not counted, only that in the online deployment stage is included in the comparison. The CPU runtime and the logarithmic rate, represented as \(\sum_{k\in[K]}\ln\left(R_k\right)\) are averaged over 50 randomly distributed samples in a 5-user scenario. We observe that the DL method performs almost the same as the AO method, but has orders of magnitude lower complexity. Therefore, in subsequent simulations, we shall only showcase the performance of the DL method.

\subsection{Array Gain with Assigned Subbands}
\begin{table}[t]
\renewcommand\arraystretch{1.1}
\centering
\caption{The array gain and logarithmic rate achieved by different beamforming architectures for a two-user scenario}
\label{Table array gain}
\begin{tabular}{|c| c| c| c|}
\hline
\textbf{Architecture}& \textbf{Array gain 1} &\textbf{Array gain 2} &\textbf{Logarithmic Rate} \\
\hline
FD & 63.97 & 63.99 & 50.12\\
\hline
PA (\(N_{\rm{T}}\!=\!0\)) & 30.48 & 37.33 & 49.73\\
\hline
JPTA (\(N_{\rm{T}}\!=\!16\)) & 29.91 & 42.32 & 49.81\\ 
\hline
JPTA (\(N_{\rm{T}}\!=\!64\)) & 38.45 & 46.90 & 49.91\\
\hline
\end{tabular}
\end{table}

\begin{figure}[t]
    \centering
    \includegraphics[width=0.95\linewidth]{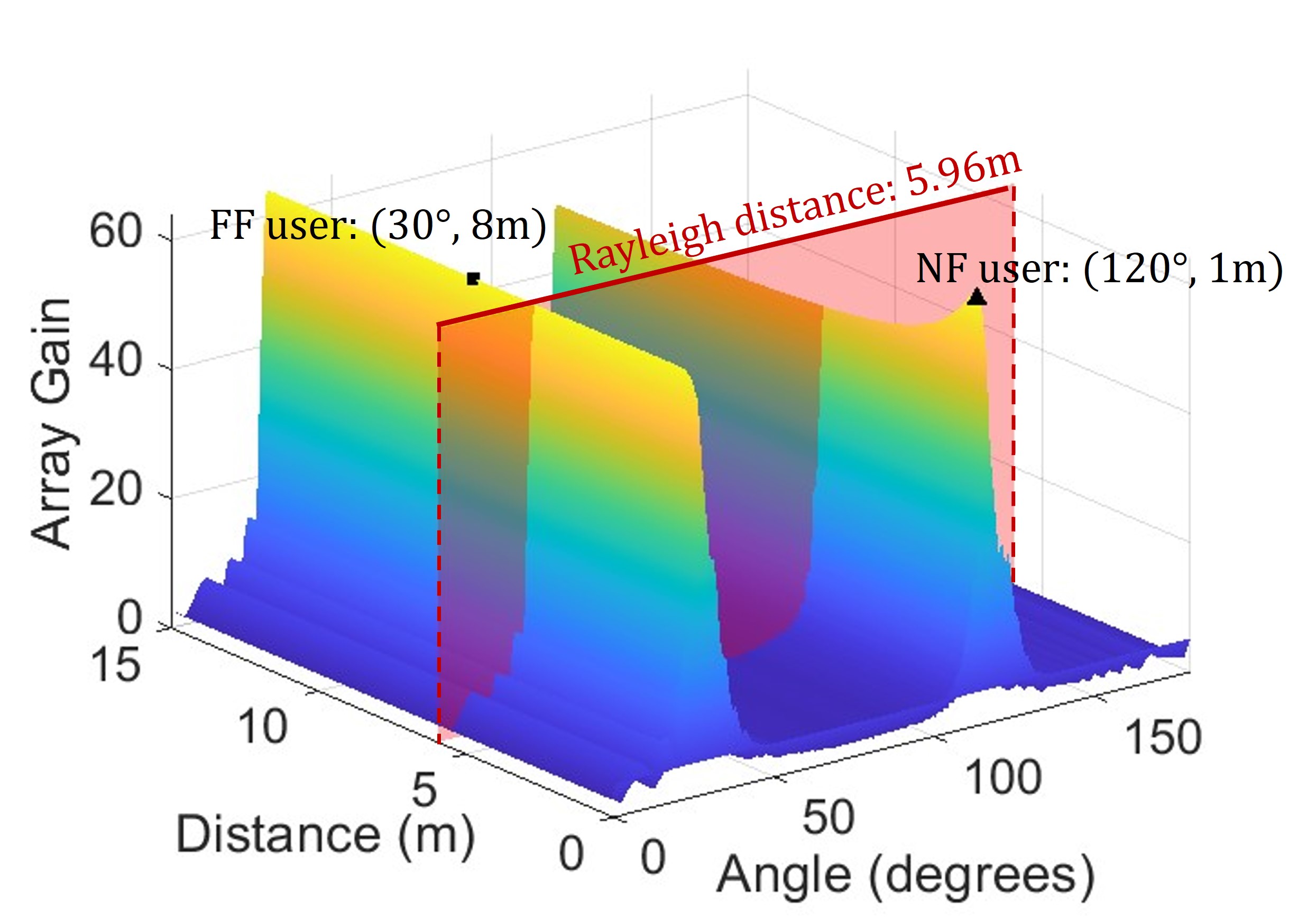}
    \caption{Average array gain of FD beamforming with assigned subbands in a two-user scenario.}
    \label{fig: NF-FF}
\end{figure}
\begin{figure}[t]
    \centering
    \subfloat[FD]{\includegraphics[width=0.48\linewidth]{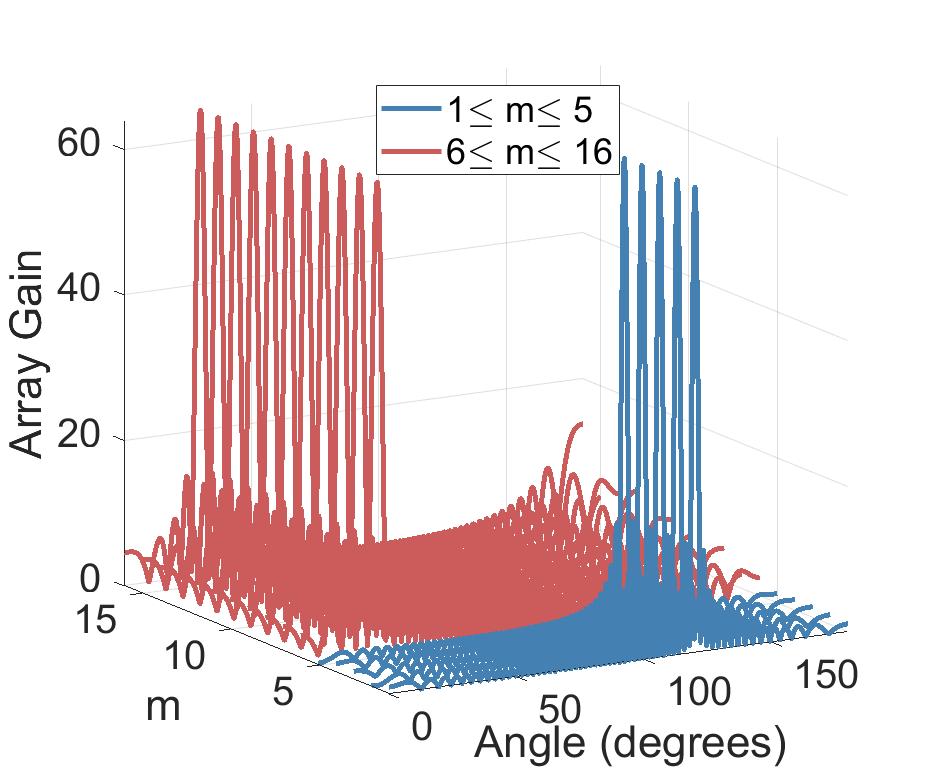}
    \label{c1}}
    \hfil
    \subfloat[PA]{\includegraphics[width=0.48\linewidth]{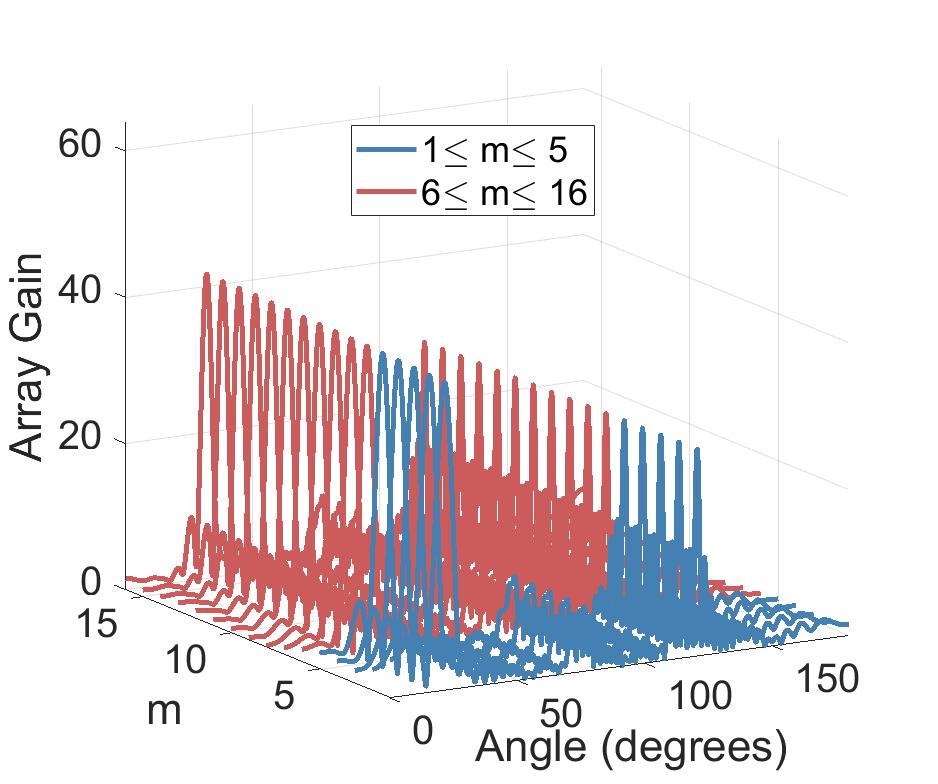}
    \label{c2}}
    \hfil
    \subfloat[JPTA (\(N_{\rm{T}}\!\!=\!\!16\))]{\includegraphics[width=0.48\linewidth]{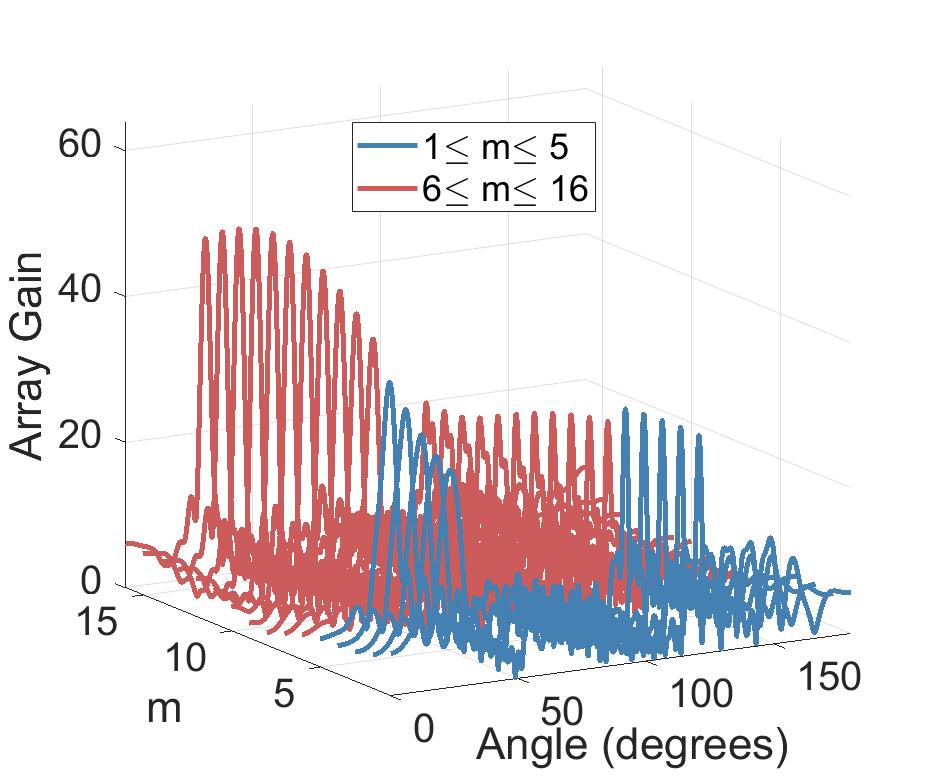}
    \label{c3}} % 
    \hfil
    \subfloat[JPTA (\(N_{\rm{T}}\!\!=\!\!64\))]{\includegraphics[width=0.48\linewidth]{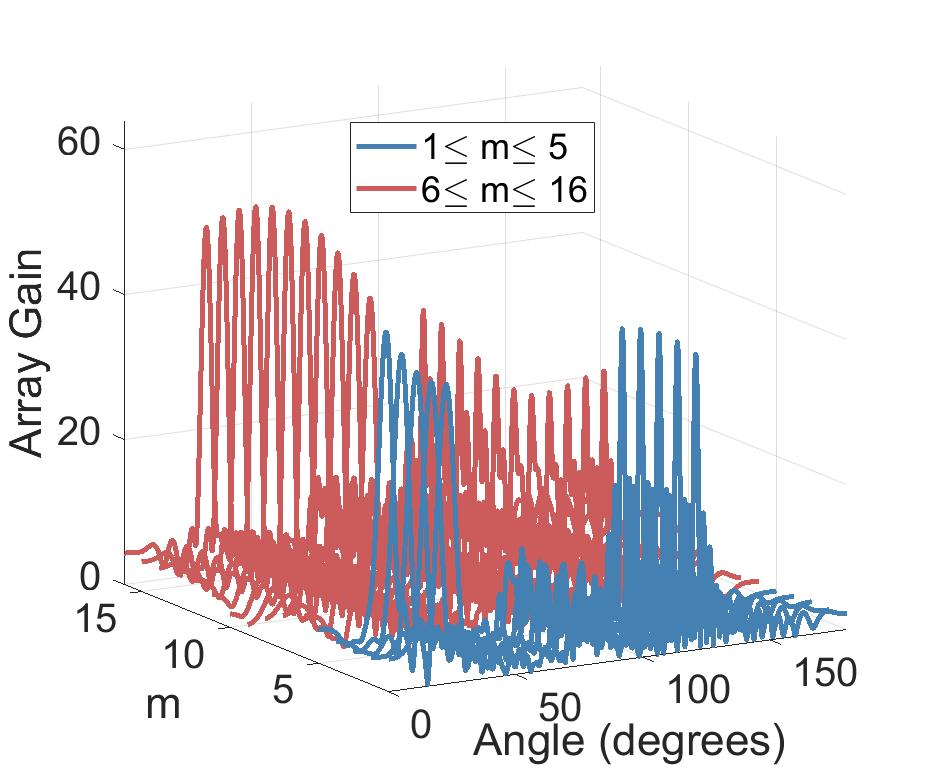}
    \label{c4}}
    \caption{Array gain for 2-user scenario at the distance of 1 meters achieved by different approaches.}
    \label{fig: gain}
\end{figure}
In this subsection, we illustrate the array gain achieved by the proposed JPTA beamforming under given subband allocation. 

To demonstrate the beam pattern distinctions between NF and FF beamforming, Fig. \ref{fig: NF-FF} depicts the average array gain over the assigned subbands in a two-user scenario (one NF user and one FF user) achieved by FD beamforming. In this scenario, subbands \(1\leq m \leq 5\) are allocated to the NF user at \((120^{\circ}, 1\) m) and subbands \(6\leq m \leq 16\) are assigned to the FF user at \((30^{\circ}, 8\) m). It can be observed that the near-field beam is focused on a specific point, whereas the far-field beam is directed towards a general direction.

Fig. \ref{fig: gain} and Table \ref{Table array gain} show the array gain for the same two-user scenario employing different antenna architectures with fixed subband allocations. The PA generates a frequency-flat response, maintaining consistent array gain across all subbands. In contrast, the JPTA produces a frequency-dependent array gain, effectively enhancing the gain at specific frequency bands for each user. It can also be observed from Fig. \ref{fig: gain} and Table \ref{Table array gain} that employing more TTD units within JPTA increases the total array gain at targeted locations, which correspondingly results in higher logarithmic rates for the users. 
% This targeted enhancement leads to significant performance improvements for both users under the same conditions. 

\subsection{Achievable User Rate}
\begin{figure}[t]
    \centering
    \hfil
    \subfloat[\(K=2\)]{\includegraphics[width=1\linewidth]{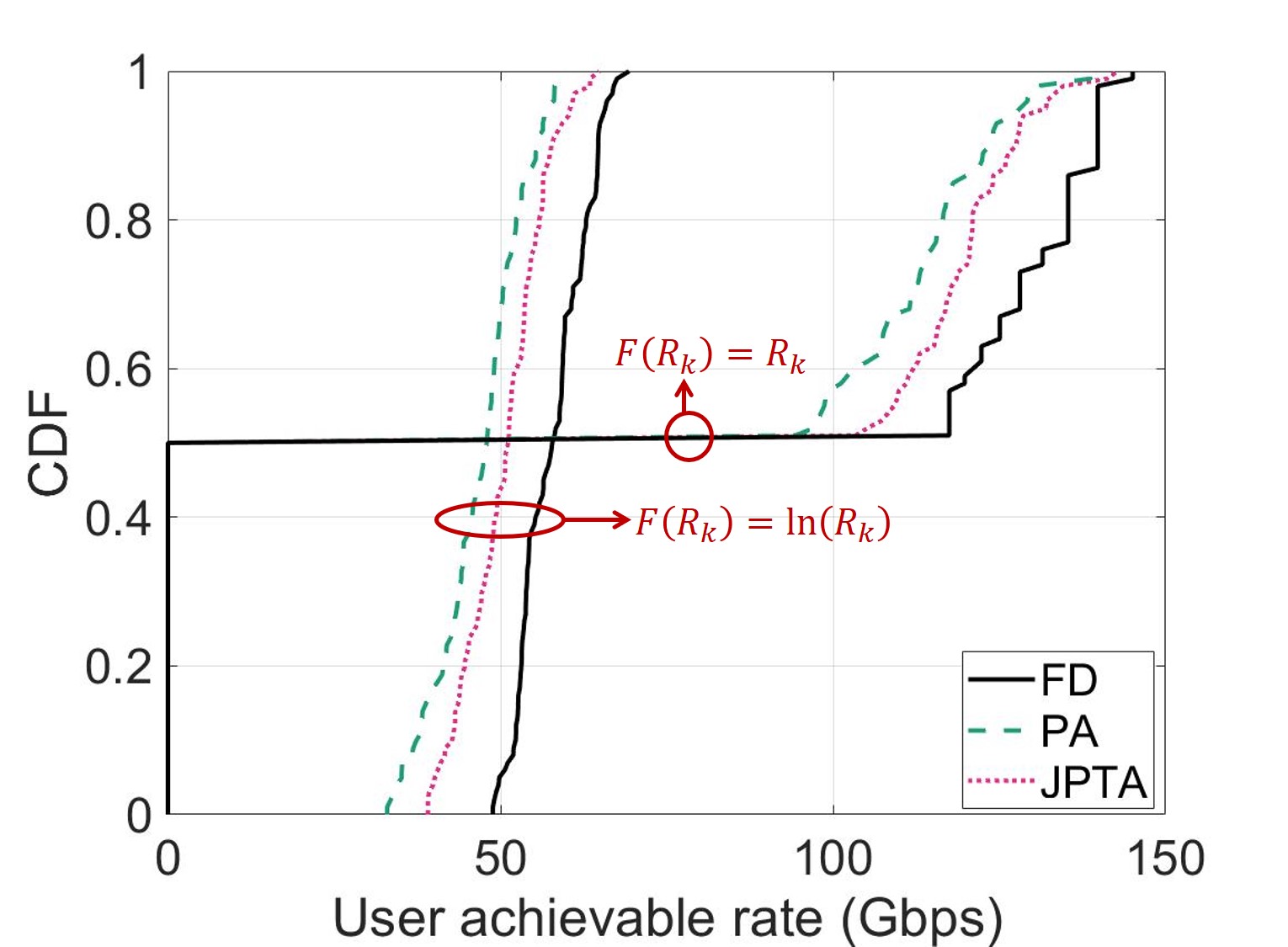}}
    \hfil
    \subfloat[\(K=5\)]{\includegraphics[width=1\linewidth]{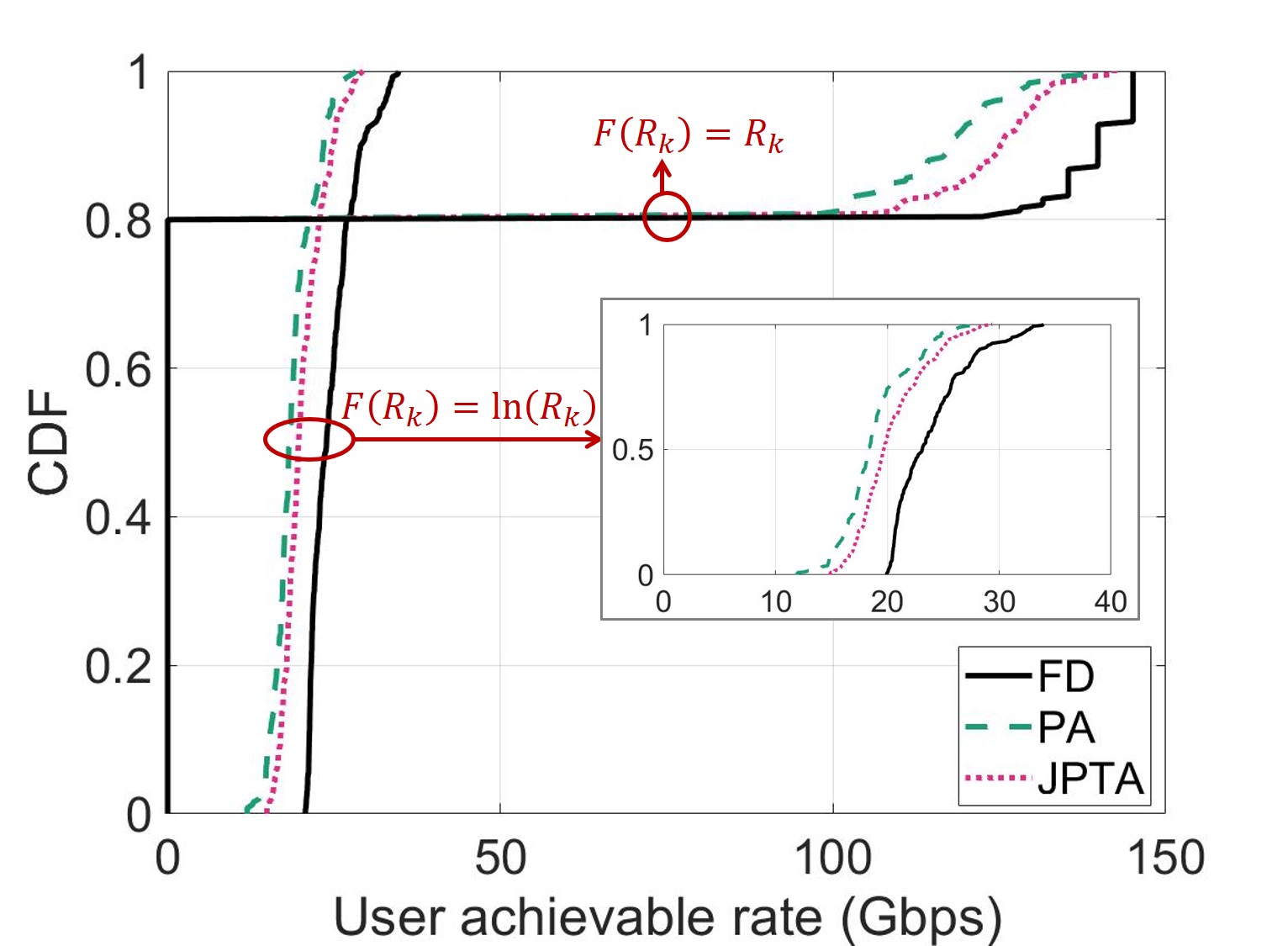}}
    \caption{CDF of user rate under different optimization goals and approaches.}
    \label{fig: CDF}
\end{figure}
This subsection compares the achievable user rates using different approaches and optimization objectives. Fig. \ref{fig: CDF} (a) and (b) show the cumulative distribution function (CDF) of user rates across 50 randomly distributed samples with \(K=2\) and \(K=5\), respectively. The results show that when adopting a logarithmic utility function (\(F(R_k)=\ln(R_k)\)), all users can be served by the BS. In contrast, optimizing for sum-rate (\(F(R_k)=R_k\)) leads to biased resource allocation, favoring users closer to the BS. It can be observed that maximizing the total logarithmic rate more effectively balances achievable rate and fairness among users at varying distances. Comparative analysis under distinct antenna architectures shows that the JPTA significantly boosts user rates and reduces the prevalence of low rates. Specifically, integrating TTDs into the beamforming architecture under a logarithmic utility function increases the average rates by 8.21\% in 2-user scenarios and by 8.07\% in 5-user scenarios. Under a sum-rate optimization, the enhancements are 6.97\% and 7.15\% respectively, illustrating the effectiveness of TTDs in mitigating the spatial-wideband effect when serving a single user. These findings underscore the JPTA's capability to significantly boost service quality for multiple users using a single RF chain. 
% Moreover, it can be observed that the performance of the DL approach is comparable to that of the AO algorithm in terms of user rates.

\subsection{SE, Logarithmic Rate, and EE Performance}
In this subsection, we evaluate the influence of bandwidth on SE and examine the effects of the number of TTDs and maximum delay range \(\tau_{\max}\) on logarithmic rate and EE.

\begin{figure}[t]
    \centering
    \includegraphics[width=1\linewidth]{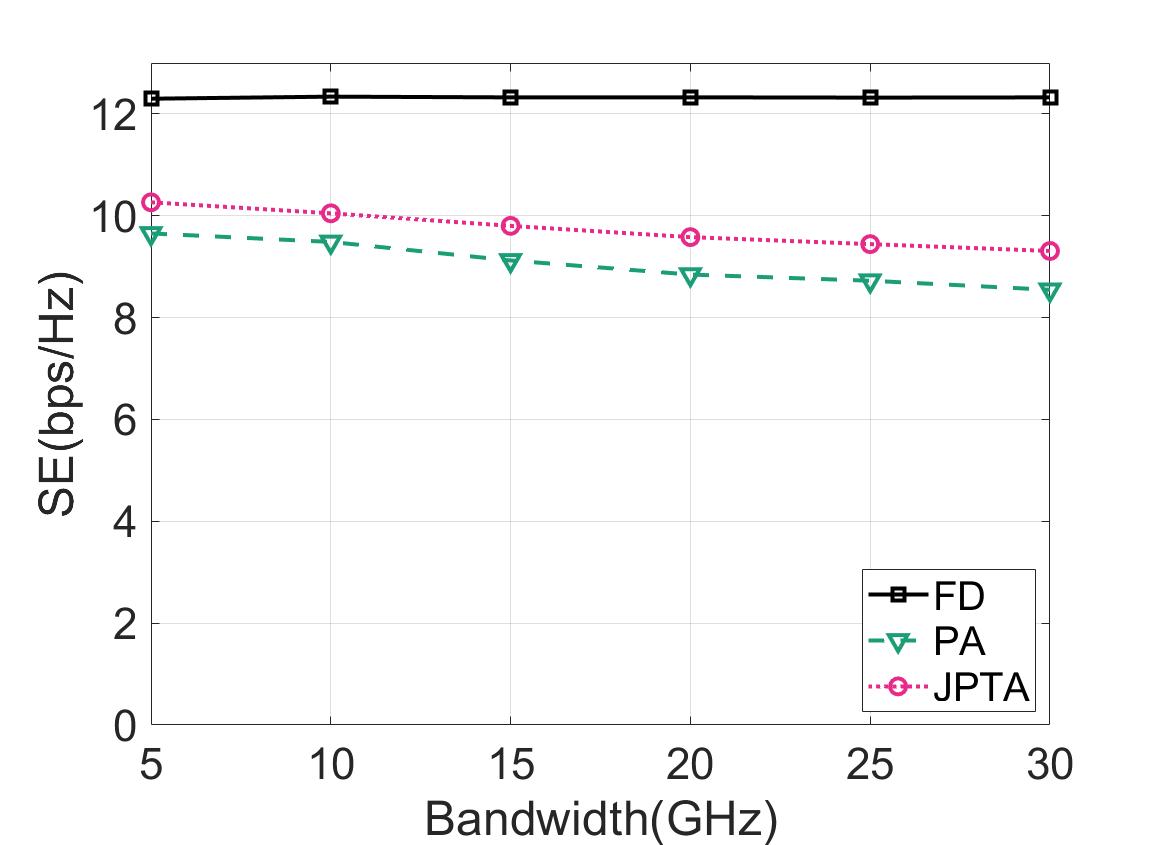}
    \caption{Average SE versus the bandwidth when \(K=5\).}
    \label{fig: BW}
\end{figure}

Fig. \ref{fig: BW} explores the impact of bandwidth on SE with a fixed signal-to-noise ratio when \(K=5\). SE, defined as the ratio of sum-rate to bandwidth, decreases with the bandwidth in both AO and DL approaches. This decrease is attributed to the wider bandwidth dispersing frequency beams more extensively, complicating the control of beam directions at different frequencies. Moreover, JPTA enhances SE by approximately 8.22\% compared to PA. 

\begin{figure}[t]
    \centering
    \includegraphics[width=1\linewidth]{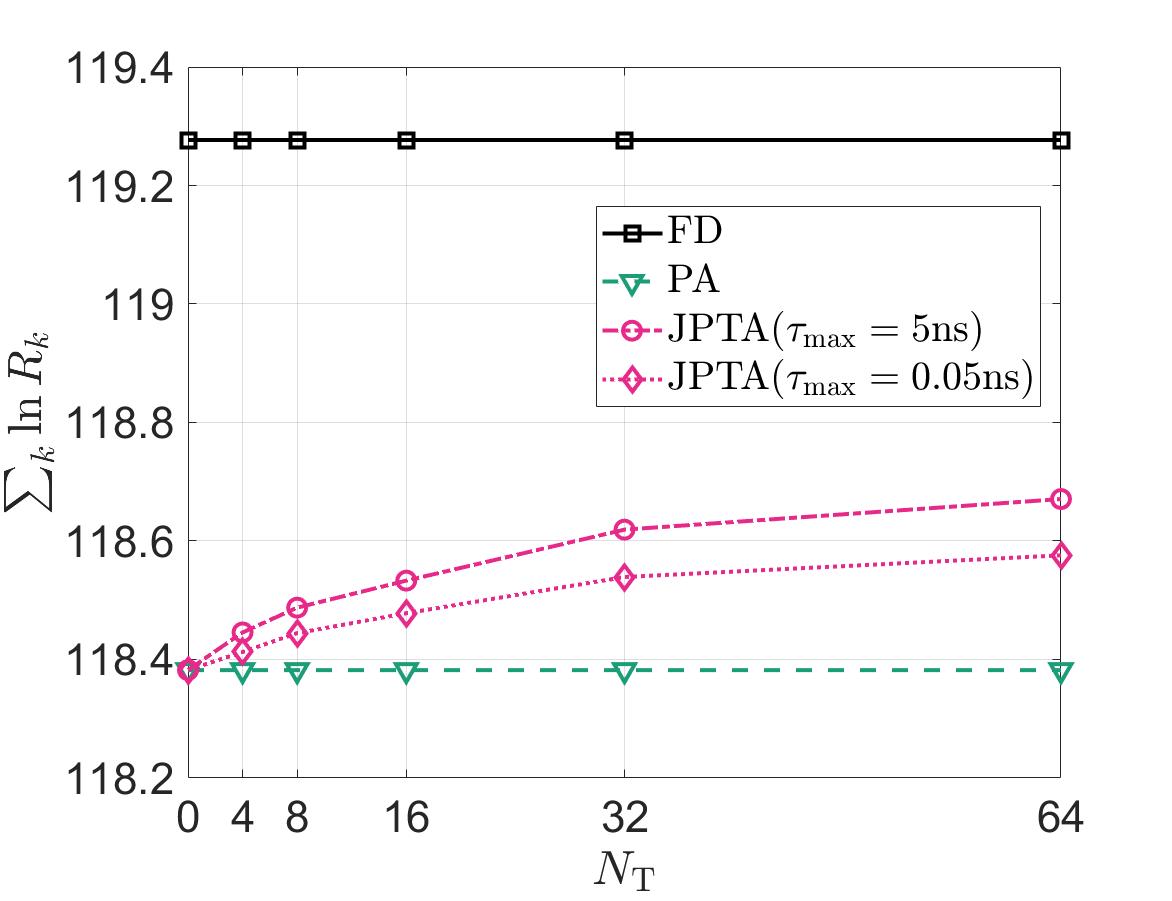}
    \caption{The impact of \(N_{\rm{T}}\) and \(\tau_{\max}\) on the average logarithmic rate when \(K=5\).}
    \label{fig: TTD}
\end{figure}
\begin{figure}[t]
    \centering
    \includegraphics[width=1\linewidth]{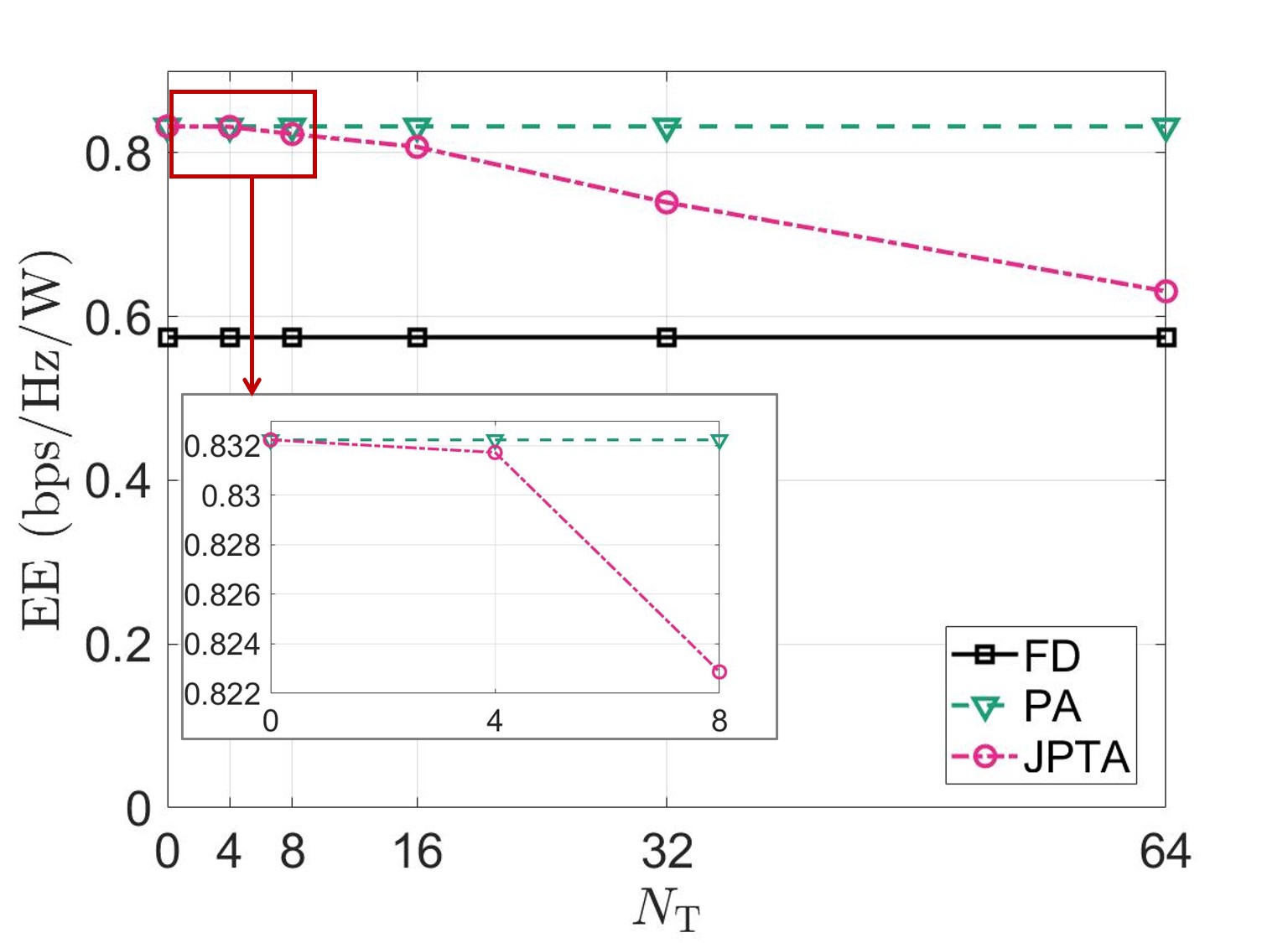}
    \caption{Average EE versus \(N_{\rm{T}}\) when \(K=5\).}
    \label{fig: EE}
\end{figure}

In Fig. \ref{fig: TTD}, we analyze the impact of varying numbers of TTDs and different maximum delay ranges on the logarithmic rate in a 5-user scenario. The figure shows that the average logarithmic rate increases with the number of TTDs. Specifically, with a maximum delay \(\tau_{\max}\) of 5 ns and each antenna connected to a TTD (\(N_{\rm{T}}=64\)), the difference in logarithmic rates between the JPTA-based and FD beamforming narrows by 26.4\% compared to the difference between PA (\(N_{\rm{T}}=0\)) and FD approaches. Reducing \(\tau_{\max}\) to 0.05 ns decreases this enhancement in logarithmic rate difference to 17.3\%. 

In Fig. \ref{fig: EE}, we examine the impact of the number of TTDs on EE. We use practical values for power consumption: $P_{\rm{BB}}=300$ mW for the digital beamformer \cite{dai2022delay}, $P_{\rm{RF}}=200$ mW for the RF chain \cite{7370753}, $P_{\rm{PS}}=30$ mW for the PS \cite{7370753}, and $P_{\rm{TTD}}=100$ mW for each TTD \cite{cho2018true}. Consequently, the power consumption for FD, PA, and JPTA beamforming can be expressed as \(P_{\rm{FD}} = P_{\rm{t}} + P_{\rm{BB}} + NP_{\rm{RF}}\), \(P_{\rm{PA}} = P_{\rm{t}} + P_{\rm{BB}} + P_{\rm{RF}} + NP_{\rm{PS}}\), and \(P_{\rm{JPTA}} = P_{\rm{t}} + P_{\rm{BB}} + P_{\rm{RF}} + N_{\rm{T}}P_{\rm{TTD}} + NP_{\rm{PS}}\), respectively. EE is defined as the ratio of SE to power consumption. As observed from Fig. \ref{fig: EE}, the PA achieves the highest EE, while the EE of the JPTA finds a middle ground between that of FD and PA, effectively balancing the two approaches. Moreover, the EE of the JPTA decreases with an increasing number of TTDs, due to the rising power demands of additional TTDs surpassing the improvements in SE.

% EE只要比FD好就行，JPTA在FD和PA中stride a balance，取到了中间
% the energy efficiency of the sub-connected architecture exhibits a continuous decline from the outset, as the installation of additional TTDs fails to provide significant improvements in spectral efficiency

\section{Conclusion}
\label{sec: conclusion}
This paper investigates resource allocation and beamforming optimization in hybrid NF and FF wideband systems using JPTA, focusing on its capacity to generate frequency-dependent beams for simultaneous multi-user service. We introduce a 3-step AO method and an innovative end-to-end unsupervised learning network to maximize the concave utility function of user rates. The AO method iteratively optimizes the subband allocation, analog beamforming, and power distribution. The DL approach integrates a feature extraction module, a graph attention module, and a normalization module to learn to map channel information to resource allocation and analog beamforming strategies. The Gumbel-softmax trick, applied within the normalization module, effectively manages the discrete constraints associated with subband allocation.

We analyze the performance of various antenna architectures and optimization objectives and examined the impacts of bandwidth, the number of TTDs, and the maximum delay range. Our numerical results demonstrate that JPTA surpasses conventional PA systems in terms of achievable rates and user fairness. JPTA also strikes a good balance between PA and FD in the term of EE. Moreover, the performance of the proposed DL approach is comparable to that of traditional optimization methods with orders of magnitude lower computational complexity, highlighting its potential as an effective and cost-efficient solution for complex beamforming challenges in next-generation wireless networks.

\bibliographystyle{IEEEtran}
\bibliography{reference}

\begin{thebibliography}{10}
\providecommand{\url}[1]{#1}
\csname url@samestyle\endcsname
\providecommand{\newblock}{\relax}
\providecommand{\bibinfo}[2]{#2}
\providecommand{\BIBentrySTDinterwordspacing}{\spaceskip=0pt\relax}
\providecommand{\BIBentryALTinterwordstretchfactor}{4}
\providecommand{\BIBentryALTinterwordspacing}{\spaceskip=\fontdimen2\font plus
\BIBentryALTinterwordstretchfactor\fontdimen3\font minus \fontdimen4\font\relax}
\providecommand{\BIBforeignlanguage}[2]{{%
\expandafter\ifx\csname l@#1\endcsname\relax
\typeout{** WARNING: IEEEtran.bst: No hyphenation pattern has been}%
\typeout{** loaded for the language `#1'. Using the pattern for}%
\typeout{** the default language instead.}%
\else
\language=\csname l@#1\endcsname
\fi
#2}}
\providecommand{\BIBdecl}{\relax}
\BIBdecl

\bibitem{wcnc}
Y.~Cai, M.~Tao, and S.~Sun, ``{Frequency-Dependent Beamforming for Hybrid Near-Far Field Communications Through Joint Phase-Time Arrays},'' \emph{accepted by IEEE Wireless Communications and Networking Conference}, 2025.

\bibitem{wang2023road}
C.-X. Wang, X.~You, X.~Gao, X.~Zhu, Z.~Li, C.~Zhang, H.~Wang, Y.~Huang, Y.~Chen, H.~Haas \emph{et~al.}, ``{On the road to 6G: Visions, requirements, key technologies, and testbeds},'' \emph{IEEE Communications Surveys \& Tutorials}, vol.~25, no.~2, pp. 905--974, 2023.

\bibitem{rappaport2019wireless}
T.~S. Rappaport, Y.~Xing, O.~Kanhere, S.~Ju, A.~Madanayake, S.~Mandal, A.~Alkhateeb, and G.~C. Trichopoulos, ``{Wireless communications and applications above 100 GHz: Opportunities and challenges for 6G and beyond},'' \emph{IEEE access}, vol.~7, pp. 78\,729--78\,757, 2019.

\bibitem{busari2017millimeter}
S.~A. Busari, K.~M.~S. Huq, S.~Mumtaz, L.~Dai, and J.~Rodriguez, ``{Millimeter-wave massive MIMO communication for future wireless systems: A survey},'' \emph{IEEE Communications Surveys \& Tutorials}, vol.~20, no.~2, pp. 836--869, 2017.

\bibitem{bjornson2019massive}
E.~Bj{\"o}rnson, L.~Sanguinetti, H.~Wymeersch, J.~Hoydis, and T.~L. Marzetta, ``{Massive MIMO is a reality—What is next?: Five promising research directions for antenna arrays},'' \emph{Digital Signal Processing}, vol.~94, pp. 3--20, 2019.

\bibitem{molisch2017hybrid}
A.~F. Molisch, V.~V. Ratnam, S.~Han, Z.~Li, S.~L.~H. Nguyen, L.~Li, and K.~Haneda, ``{Hybrid beamforming for massive MIMO: A survey},'' \emph{IEEE Communications magazine}, vol.~55, no.~9, pp. 134--141, 2017.

\bibitem{el2014spatially}
O.~El~Ayach, S.~Rajagopal, S.~Abu-Surra, Z.~Pi, and R.~W. Heath, ``{Spatially sparse precoding in millimeter wave MIMO systems},'' \emph{IEEE transactions on wireless communications}, vol.~13, no.~3, pp. 1499--1513, 2014.

\bibitem{rotman2016true}
R.~Rotman, M.~Tur, and L.~Yaron, ``{True time delay in phased arrays},'' \emph{Proceedings of the IEEE}, vol. 104, no.~3, pp. 504--518, 2016.

\bibitem{liu2023near1}
Y.~Liu, Z.~Wang, J.~Xu, C.~Ouyang, X.~Mu, and R.~Schober, ``{Near-field communications: A tutorial review},'' \emph{IEEE Open Journal of the Communications Society}, 2023.

\bibitem{liu2023near}
Y.~Liu, J.~Xu, Z.~Wang, X.~Mu, and L.~Hanzo, ``{Near-field communications: What will be different?}'' \emph{arXiv preprint arXiv:2303.04003}, 2023.

\bibitem{zhang20236g}
H.~Zhang, N.~Shlezinger, F.~Guidi, D.~Dardari, and Y.~C. Eldar, ``{6G wireless communications: From far-field beam steering to near-field beam focusing},'' \emph{IEEE Communications Magazine}, vol.~61, no.~4, pp. 72--77, 2023.

\bibitem{dai2022delay}
L.~Dai, J.~Tan, Z.~Chen, and H.~V. Poor, ``{Delay-phase precoding for wideband THz massive MIMO},'' \emph{IEEE Transactions on Wireless Communications}, vol.~21, no.~9, pp. 7271--7286, 2022.

\bibitem{ratnam2022joint}
V.~V. Ratnam, J.~Mo, A.~Alammouri, B.~L. Ng, J.~Zhang, and A.~F. Molisch, ``{Joint phase-time arrays: A paradigm for frequency-dependent analog beamforming in 6G},'' \emph{IEEE Access}, vol.~10, pp. 73\,364--73\,377, 2022.

\bibitem{yildiz20243d}
O.~Yildiz, A.~AlAmmouri, J.~Mo, Y.~Nam, E.~Erkip \emph{et~al.}, ``{3D Beamforming Through Joint Phase-Time Arrays},'' \emph{arXiv preprint arXiv:2401.00819}, 2024.

\bibitem{alammouri2022extending}
A.~Alammouri, J.~Mo, V.~V. Ratnam, B.~L. Ng, R.~W. Heath, J.~Lee, and J.~Zhang, ``{Extending uplink coverage of mmwave and terahertz systems through joint phase-time arrays},'' \emph{IEEE Access}, vol.~10, pp. 88\,872--88\,884, 2022.

\bibitem{jain2023mmflexible}
I.~K. Jain, R.~R. Vennam, R.~Subbaraman, and D.~Bharadia, ``{mmflexible: Flexible directional frequency multiplexing for multi-user mmwave networks},'' in \emph{IEEE INFOCOM 2023-IEEE Conference on Computer Communications}.\hskip 1em plus 0.5em minus 0.4em\relax IEEE, 2023, pp. 1--10.

\bibitem{zhao2024fast}
D.~Zhao, I.~Pehlivan, A.~Wadaskar, and D.~Cabric, ``{Fast Frequency-Direction Mapping Design for Data Communication With True-Time-Delay Array Architecture},'' in \emph{2024 International Conference on Computing, Networking and Communications (ICNC)}.\hskip 1em plus 0.5em minus 0.4em\relax IEEE, 2024, pp. 1071--1076.

\bibitem{10179244}
S.~Kim, J.~Park, J.~Moon, and B.~Shim, ``{Fast and Accurate Terahertz Beam Management via Frequency-Dependent Beamforming},'' \emph{IEEE Transactions on Wireless Communications}, vol.~23, no.~3, pp. 1699--1712, 2024.

\bibitem{gao2023integrated}
F.~Gao, L.~Xu, and S.~Ma, ``{Integrated sensing and communications with joint beam-squint and beam-split for mmWave/THz massive MIMO},'' \emph{IEEE Transactions on Communications}, vol.~71, no.~5, pp. 2963--2976, 2023.

\bibitem{10271123}
H.~Luo, F.~Gao, W.~Yuan, and S.~Zhang, ``{Beam Squint Assisted User Localization in Near-Field Integrated Sensing and Communications Systems},'' \emph{IEEE Transactions on Wireless Communications}, vol.~23, no.~5, pp. 4504--4517, 2024.

\bibitem{cui2022near}
M.~Cui, L.~Dai, Z.~Wang, S.~Zhou, and N.~Ge, ``{Near-field rainbow: Wideband beam training for XL-MIMO},'' \emph{IEEE Transactions on Wireless Communications}, vol.~22, no.~6, pp. 3899--3912, 2022.

\bibitem{10541333}
M.~Cui and L.~Dai, ``{Near-Field Wideband Beamforming for Extremely Large Antenna Arrays},'' \emph{IEEE Transactions on Wireless Communications}, pp. 1--1, 2024.

\bibitem{guo2024wideband}
Y.~Guo, Y.~Zhang, Z.~Wang, and Y.~Liu, ``{Wideband Beamforming for Near-Field Communications with Circular Arrays},'' \emph{arXiv preprint arXiv:2404.02811}, 2024.

\bibitem{10458958}
Z.~Wang, X.~Mu, Y.~Liu, and R.~Schober, ``{TTD Configurations for Near-Field Beamforming: Parallel, Serial, or Hybrid?}'' \emph{IEEE Transactions on Communications}, vol.~72, no.~6, pp. 3783--3799, 2024.

\bibitem{ting2024adaptive}
H.~Ting, Z.~Wang, and Y.~Liu, ``{Adaptive TTD Configurations for Near-Field Communications: An Unsupervised Transformer Approach},'' \emph{arXiv preprint arXiv:2403.18146}, 2024.

\bibitem{zhang2023deep}
Y.~Zhang and A.~Alkhateeb, ``{Deep learning of near field beam focusing in terahertz wideband massive MIMO systems},'' \emph{IEEE Wireless Communications Letters}, vol.~12, no.~3, pp. 535--539, 2023.

\bibitem{lin2019beamforming}
T.~Lin and Y.~Zhu, ``{Beamforming design for large-scale antenna arrays using deep learning},'' \emph{IEEE Wireless Communications Letters}, vol.~9, no.~1, pp. 103--107, 2019.

\bibitem{liu2022deep}
Z.~Liu, Y.~Yang, F.~Gao, T.~Zhou, and H.~Ma, ``{Deep unsupervised learning for joint antenna selection and hybrid beamforming},'' \emph{IEEE Transactions on Communications}, vol.~70, no.~3, pp. 1697--1710, 2022.

\bibitem{shen2022graph}
Y.~Shen, J.~Zhang, S.~Song, and K.~B. Letaief, ``{Graph neural networks for wireless communications: From theory to practice},'' \emph{IEEE Transactions on Wireless Communications}, vol.~22, no.~5, pp. 3554--3569, 2022.

\bibitem{li2024gnn}
Y.~Li, Y.~Lu, B.~Ai, O.~A. Dobre, Z.~Ding, and D.~Niyato, ``{GNN-based beamforming for sum-rate maximization in MU-MISO networks},'' \emph{IEEE Transactions on Wireless Communications}, 2024.

\bibitem{veličković2018graph}
P.~Veličković, G.~Cucurull, A.~Casanova, A.~Romero, P.~Liò, and Y.~Bengio, ``{Graph Attention Networks},'' in \emph{International Conference on Learning Representations}, 2018.

\bibitem{liu2024near}
Y.~Liu, C.~Ouyang, Z.~Wang, J.~Xu, X.~Mu, and A.~L. Swindlehurst, ``{Near-field communications: A comprehensive survey},'' \emph{arXiv preprint arXiv:2401.05900}, 2024.

\bibitem{kelly1997charging}
F.~Kelly, ``{Charging and rate control for elastic traffic},'' \emph{European Transactions on Telecommunications}, vol.~8, no.~1, pp. 33--37, 1997.

\bibitem{grant2014cvx}
M.~Grant and S.~Boyd, ``{CVX: Matlab software for disciplined convex programming, version 2.1},'' 2014.

\bibitem{boyd2004convex}
S.~Boyd and L.~Vandenberghe, \emph{{Convex optimization}}.\hskip 1em plus 0.5em minus 0.4em\relax Cambridge university press, 2004.

\bibitem{goldsmith2005wireless}
A.~Goldsmith, \emph{{Wireless communications}}.\hskip 1em plus 0.5em minus 0.4em\relax Cambridge university press, 2005.

\bibitem{jang2017categorical}
E.~Jang, S.~Gu, and B.~Poole, ``{Categorical Reparameterization with Gumbel-Softmax},'' in \emph{International Conference on Learning Representations}, 2017.

\bibitem{7370753}
R.~Méndez-Rial, C.~Rusu, N.~González-Prelcic, A.~Alkhateeb, and R.~W. Heath, ``{Hybrid MIMO Architectures for Millimeter Wave Communications: Phase Shifters or Switches?}'' \emph{IEEE Access}, vol.~4, pp. 247--267, 2016.

\bibitem{cho2018true}
M.-K. Cho, I.~Song, and J.~D. Cressler, ``{A true time delay-based SiGe bi-directional T/R chipset for large-scale wideband timed array antennas},'' in \emph{2018 IEEE Radio Frequency Integrated Circuits Symposium (RFIC)}.\hskip 1em plus 0.5em minus 0.4em\relax IEEE, 2018, pp. 272--275.

\end{thebibliography}
\end{document}